\documentclass[bibyear]{aa} 

\usepackage{graphics}
\usepackage{graphicx}
\usepackage{epstopdf}
\usepackage{pdfpages}
\usepackage{txfonts}
\begin{document}

   \title{The XXL Survey }

   \subtitle{XXVI. Optical and near-infrared identifications of the ATCA 2.1 GHz radio sources  in the XXL-S Field }

   \author{Paolo Ciliegi\inst{1},  Nika Jurlin\inst{2}\thanks{moved to Kapteyn Astronomical Institute, University of Groningen, Postbus 800, NL-9700 AV Groningen, the Netherlands  ~~and to 
    ASTRON, the Netherlands Institute for Radio Astronomy, Postbus 2, 7990 AA, Dwingeloo, The Netherlands } , Andrew Butler\inst{3} , Jacinta Delhaize\inst{2}, Sotiria Fotopoulou\inst{4}, Minh Huynh\inst{3,5}, Angela Iovino\inst{6}, Vernesa   Smol{\v c}i{\'c}\inst{2},   Lucio Chiappetti\inst{7}  \and Marguerite Pierre\inst{8,9} 
          }

   \institute{$^1$INAF- Osservatorio di Astrofisica e Scienza dello Spazio di Bologna, via Gobetti 93/3,   40129, Bologna, Italy      \email{paolo.ciliegi@inaf.it} \\
             $^2$University of Zagreb, Physics Department, Bijenicka cesta 32, 10002 Zagreb,  Croatia\\
             $^3$International Centre for Radio Astronomy Research (ICRAR), University of Western Australia, 35 Stirling Hwy, Crawley WA 6009, Australia \\
             $^4$Centre for Extragalactic Astronomy, Department of Physics, Durham University, South Road, Durham, DH1 3LE, UK\\             
             $^5$CSIRO Astronomy and Space Science, 26 Dick Perry Ave, Kensington WA 6151, Australia \\ 
             $^6$INAF - Osservatorio Astronomico di Brera, via Brera 28,  20121 Milano, Italy \\ 
             $^7$INAF -  IASF Milano, via Bassini 15, 20133 Milano, Italy \\
             $^8$IRFU, CEA, Universit\'e Paris-Saclay, F-91191 Gif sur Yvette, France\\
             $^9$Universit\'e Paris Diderot, AIM, Sorbonne Paris Cit\'e, CEA, CNRS, F-91191 Gif sur Yvette, France 
             }

       \date{Received date / accepted date}
       
 \abstract{}{}{}{}
 
  \abstract
    {To investigate the nature of the extragalactic radio sources it is  necessary to couple radio 
      surveys with multiwavelength  observations over large areas of the sky.   The XMM-Newton Extragalactic (XXL) 
      survey is the largest survey ever undertaken with the XMM-Newton X-ray  telescope over two separate fields of 25 deg$^{2}$ 
      each (XXL-N and XXL-S).   At the same time the XXL survey benefits from a wealth of ancillary 
      data spanning from the far-ultraviolet to the  mid-infrared.}
      { In this paper we present the optical, near-infrared (NIR) and X-ray identifications of the 6287 radio
        sources detected in the 2.1 GHz deep radio survey down to a median rms of $\sigma\approx$ 41$\mu$Jy/beam
        obtained with the Australia Telescope Compact Array (ATCA) in the XXL-S field. 
         The goal of this paper is to provide a multiwavelength catalogue of the counterparts of the radio sources to be 
        used in further studies.}
       {For the optical and NIR identification of the radio sources, we used the likelihood ratio (LR) technique, slightly modified in order to take into account 
        the presence of a large number of relatively bright  counterparts close to the radio sources. The LR technique was applied to seven 
        optical bands ($g_{BCS}$, $g_{DEC}$, $r_{BCS}$, $r_{DEC}$, $i_{BCS}$, $i_{DEC}$, $z_{DEC}$) and to three NIR bands ($J$, $H$, $K$).   
         }
       {The ten different photometric catalogues  have been combined into a single master catalogue where all the photometric 
       information in the optical, NIR, and X-ray bands  have been collected for the counterparts of the radio sources.  This 
       procedure led to the identification of  optical/NIR counterparts for 4770 different radio sources ($\sim$77\% of the whole radio sample),   
       414 of which also have  an X-ray counterpart. This fraction of identification  is in agreement with  previous radio-optical association studies at a similar optical magnitude depth, but
        is relatively low in comparison to recent  work conducted in other radio fields using deeper optical and NIR data.       }
{ The analysis of optical and NIR properties of radio sources shows that, 
        regardless of the radio flux limit of a radio survey, the nature of the identified sources  is strongly dependent   on the depth of the optical/NIR used in the identification 
process.  Only with deep enough optical/NIR data  will we be able to identify a significant fraction of radio sources with red ($z_{DEC}$-K)  counterparts whose radio emission  is dominated by nuclear activity rather than starburst activity. }  
 

\titlerunning{The XXL Survey :  Identifications of 2.1 GHz radio sources in the XXL-S field} 
\authorrunning{Ciliegi et al.} 

\maketitle

%

\section{Introduction}

Over the past few decades, deep radio surveys down to flux levels of a few $\mu$Jy 
confirmed the well-known rapid increase in the number of faint sources towards lower flux level. This corresponds
to an upturn in the radio source counts below a few mJy.  (Condon 1984, Windhorst et al. 1985,  
Danese et al. 1987, White et al. 1997, Richards 2000, Bondi et al. 2003, Huynh et al. 2005, Fomalont et al. 2006). 
Star formation and active galactic nuclei (AGN) are the major known processes that can create radio emission detected from 
extragalactic radio sources (Condon 1992, Condon et al. 2002), with  a minor contribution  from the supernovae remnants (Dubner and Giacani 2015).

However, on the basis of the radio emission alone it is  not
possible to distinguish between  star-forming galaxies and AGN and to reveal the nature of the radio sources. 
It has been shown that radio properties such as the distribution of monochromatic radio luminosity 
of  star-forming and AGN galaxies are comparable and overlapping (Sadler et al. 1999).  
The most straightforward technique used to determine the nature of radio sources is through
spectroscopic studies of the optical counterparts. 

During the last two decades, despite a great deal of dedicated effort 
(Hammer et al. 1995, Gruppioni et. al 1999, Prandoni et al. 2001,  Afonso et al. 2005, Fomalont et al. 2006, Padovani et al. 2007), the relative
fraction of the various populations responsible for  the sub-mJy radio counts (AGN, starburst, late- and
early-type galaxies), was not well established.     For example, while 
Fomalont et al. (2006) suggested a fraction of AGN near 40$\%$ in the sub-mJy regime,  Padovani et al. (2007) 
suggested that this fraction could be 60--80$\%$, while other authors suggested a roughly equal fraction of
 AGN and star-forming galaxies (Padovani et al. 2009).  The main reason for this unclear 
picture was that the spectroscopic work needed in the optical identification process is very demanding both 
in terms of telescope time and spectral analysis. 

In order to bypass this problem,  several authors have developed new classification methods to separate star-forming galaxies and AGN that are not based on the availability of optical spectra.   In particular Smol{\v c}i{\'c} et al. (2008) based their technique on the photometric 
rest-frame colours to classify local galaxy samples, while Bardelli et al. (2010) used the star formation rates estimated from 
the optical spectral energy distribution with those based on the radio luminosity to divide the radio sources in 
passive-AGN (i.e. with very low specific star formation), non-passive AGN, and star-forming galaxies.      The application of the photometric rest-frame colour
method to the VLA-COSMOS sources (Smol{\v c}i{\'c} et al. 2008) showed that the sub-mJy radio population is a mixture 
of roughly 30--40$\%$ of star-forming galaxies and 50--60$\%$ of AGN galaxies, with a minor contribution ($\sim 10\%$) 
of quasars.  Finally, using deeper 3 GHz radio  data in the COSMOS field,  Smol{\v c}i{\'c} et al. (2017) showed that below $\sim$100 $\mu$Jy 
the fraction of star-forming galaxies increases to $\sim$ 60\%, followed by  moderate to high radiative luminosity
AGN ($20\%$)  and  low to moderate radiative luminosity AGN ($20\%$).  The photometry was crucial for the identification of the nature of the host galaxies of the radio sources.

To investigate the nature of the sub-mJy population, it is  necessary to couple radio surveys with multiwavelength observations over a large area of the sky.   
During the last decade,  the Cosmic Evolution Survey (COSMOS, Scoville et al. 2007) has been an exceptional laboratory from this point of view. 
A deep radio VLA Survey at 1.4 GHz down to an rms value of $\sim$ 11 $\mu$Jy/beam (Schinnerer  et al. 2007, 2010)   is, in fact,  available 
over an area of  2 deg$^2$  also covered  by deep observations (from the 
radio to the X-ray) obtained with most of the major space-based (Hubble, Spitzer, GALEX, XMM-Newton, Chandra, Herschel, NuStar) 
and ground-based telescopes (e.g. Keck, Subaru, VLA, ESO-VLT, CFHT UKIRT, NOAO).   
The combination of a deep radio survey with a set of multiwavelength data has produced  milestone results in 
many different fields such as  high-redshift radio galaxies (Carilli et al. 2007), radio source counts (Bondi et al. 2008), the   nature of  the faint radio population (Smol{\v c}i{\'c} et al. 2008), the cosmic evolution of AGN (Smol{\v c}i{\'c} et al. 2009a), the radio-derived star formation rate density (Smol{\v c}i{\'c} et al. 2009b), and the  environment of radio-emitting galaxies (Bardelli et al. 2010).    Moreover,  new radio observations at 3 GHz over the entire COSMOS field down to an rms value of 2.3 $\mu$Jy/beam have been recently obtained and fully analysed  (Smol{\v c}i{\'c} et al. 2017a, Smol{\v c}i{\'c} et al. 2017b,  Delvecchio et al. 2017,  Delhaize et al. 2017,  Novak  et al. 2017). 

However, while extragalactic multiwavelength deep surveys over a relatively small area (few deg$^2$) like the COSMOS survey are the ideal tool for obtaining statistically significant samples of faint galaxies, star-forming galaxies, and AGN  in order  to study their formation and evolution,   larger area surveys (tens or hundreds of deg$^2$) are need to reduce the cosmic variance and to study large-scale structure and the environment, and to search for rare objects. 

The Ultimate $XMM-Newton$ Extragalactic  (XXL) X-ray survey (Pierre et al. 2016; hereafter XXL Paper I) is the largest survey ever undertaken with 
the XMM$-$Newton X-ray telescope over two separate fields of 25 deg$^2$ each (XXL-N and XXL-S).  Moreover, 
the XXL fields benefit from ancillary photometric observations ranging from the ultraviolet to the radio bands. 
A summary of all observations available at all wavelengths targeting the XXL fields is available in XXL Paper I. 
This new panchromatic deep survey gives us the opportunity to study the multiwavelength properties of the radio sources over a very large area, reducing  cosmic variance and opening the possibility  of 
discovering   rare 
objects and phenomena. 

In this paper we  use only the data set available in the XXL-S field (RA=23$^{h}$30$^{m}$00$^{s}$, 
DEC=$-$55$^{\circ}$00$^{\prime}$00$^{\prime\prime}$).    The aim of this paper is to present the master catalogue 
of the optical, near-infrared (NIR)  and X-ray  counterparts of the radio sources present in the XXL ATCA 2.1 GHz radio source catalogue (Butler et al. 2017; hereafter XXL Paper XVIII ) and to describe the technique used during the identification process.  A detailed description of the classification of the radio sources and  their radio luminosity,  stellar mass, and star formation rate distributions are reported in 
Butler et al. (2018;  hereafter XXL Paper XXXI), while the radio spectral indices analysis and the radio source counts are reported in 
XXL Paper XVIII.  
 
The paper  is structured as follows.  In Section 2 we briefly describe the available data set in the XXL-S field  used in this paper. 
In Section 3 we describe the technique used to associate the optical counterparts with the radio sources,  while in Section 4 we report the results obtained.   Finally, in  Section 5 we give a short 
description of the optical properties of the radio sources, and in Section 6 we  summarise our results.

\section{XXL-S field data}

\subsection{Radio data}

The 25 deg$^2$ of the  XXL-S field has been observed in the radio band  with the Australia Telescope Compact Array (ATCA) at 2.1 GHz with a resolution of $\sim$4.76$^{\prime\prime}$ and a sensitivity  of $\sigma\sim$41 $\mu$Jy/beam 
(XXL Paper XVIII; see also Smol{\v c}i{\'c} et al. 2016, hereafter XXL Paper XI).  Currently this is  the largest area radio survey probed down to this flux density level.  
A 5 $\sigma$ source catalogue of 6287 radio sources has been extracted, 48 of which are considered as multiple sources, i.e. 
 with at least two separate components.  A detailed description of the radio observations, data reduction, source extraction,  and radio source  counts is reported in XXL Paper XVIII. 

\subsection{Optical and near-infrared data }

A detailed description of the multiwavelength data available in the XXL-S field is reported in XXL Paper I and 
in Fotopoulou et al. (2016, hereafter XXL Paper  VI).  Here we describe briefly the data set used in this work.   All magnitudes are expressed  in the AB system and we quote   
the limiting magnitude as  the third quartile of the respective magnitude distribution (XXL Paper VI). 

In the optical bands, the XXL-S field is covered by the Blanco Cosmological Survey (BCS, Desai et al. 2012) in the $g,r,i,z$
bands down to a limiting magnitude of  24.14, 24.06, 23.23, and 21.68, respectively.  The BCS survey  was 
obtained using  the MOSAIC II imager at the Cerro Tololo Inter-American Observatory (CTIO).   The successor 
of this camera, the Dark Energy Camera  (DECam) has been used by the XXL collaboration (PI: C Lidman) 
to cover again the XXL-S in the $g,r,i,z$ bands, but at deeper depth than BCS,  reaching a magnitude limit of 
$g$=25.73, $r$=25.78, $i$=25.6, and $z$=24.87, respectively. 

In the NIR bands, the XXL-S field is covered by the Vista Hemisphere Survey (VHS, PI R., McMahon) a public European Southern Observatory  (ESO) large programme survey.  The field is covered down to a measured limiting magnitude of  J=21.10, H=20.77, and 
K=20.34.    

Given the strong inhomogeneity of the methods used to obtain the public  catalogues and  in order to optimise the  information obtainable
 from the available  observations, a new photometric extraction has been performed within the XXL collaboration
 (XXL Paper VI) and a multiwavelength catalogue has been obtained. This catalogue contains more than four million entries 
and  15 different magnitudes from the  ultraviolet (Galaxy Evolution Explorer, GALEX bands, Morrissey et al. 2005, Martin and GALEX Team 2005),  optical ($g,r,i,z$, from BCS and DECam surveys),   NIR (J,H,K from VISTA ), and mid-infrared (IRAC and WISE bands from 3.4$\mu$ to 22 $\mu$m,  Ashby et al. 2013, Wright et al. 2010).   A detailed description of the technique 
used to construct the  multiwavelength  catalogue is reported in XXL Paper VI. 
This multiwavelength  catalogue has been used as input in the identification process of the  XXL-S radio sources.  

\subsection{X-ray}

The {\it XMM-Newton} X-ray Telescope observed the XXL-S field as part of the {\it XMM-Newton} Extragalactic  Survey with a 
depth of $\sim$6 $\times$ 10$^{-15}$ ergs s$^{-1}$ cm$^{-2}$ in the 0.5--2 keV band and $\sim$2 $\times$ $^{-14}$ ergs s$^{-1}$ cm$^{-2}$ in the 2--10 keV band 
(XXL Paper I).   The optical and NIR identification of the X-ray sources are reported in Chiappetti et al. (2018; XXL Paper XXVII ).

\section{Likelihood technique}

\subsection{Definition}

For the optical identification of the 6287 radio sources in the XXL-S field, we
used the likelihood ratio (LR) technique (Sutherland \& Saunders 1992; Ciliegi et al. 2003; Brusa et al. 2007).  
The likelihood ratio is defined as the ratio of the probability that a given source is the true optical counterpart to the probability
that the same source is an unrelated background object, \\
\begin{equation}
LR = \frac{q(m)f(r)}{n(m)},
\end{equation}
where $q(m)$ is the expected probability distribution of the true optical counterparts as a function of magnitude, $f(r)$ is the probability distribution
function of the positional errors, and $n(m)$ is the surface density of background objects with magnitude $m$ [see Ciliegi et al. 2003 for a detailed 
discussion on the procedure to calculate $q(m)$, $f(r),$ and $n(m)$]. \\
For each source we adopted an elliptical Gaussian distribution for the positional errors with the errors in
RA and DEC on the radio position reported in the radio catalogue and assuming an  optical position uncertainty as a value of 0.3 arcsec in RA and DEC. \\
We adopted a 3\arcsec\ radius for the estimate of $q(m)$, obtained by subtracting the expected number of background objects [$n(m)$] from the observed 
total number of objects listed in the catalogue around the positions of the radio sources. \\ 

\subsection{Background source distribution correction}

\begin{figure} 
        \includegraphics[width=\columnwidth]{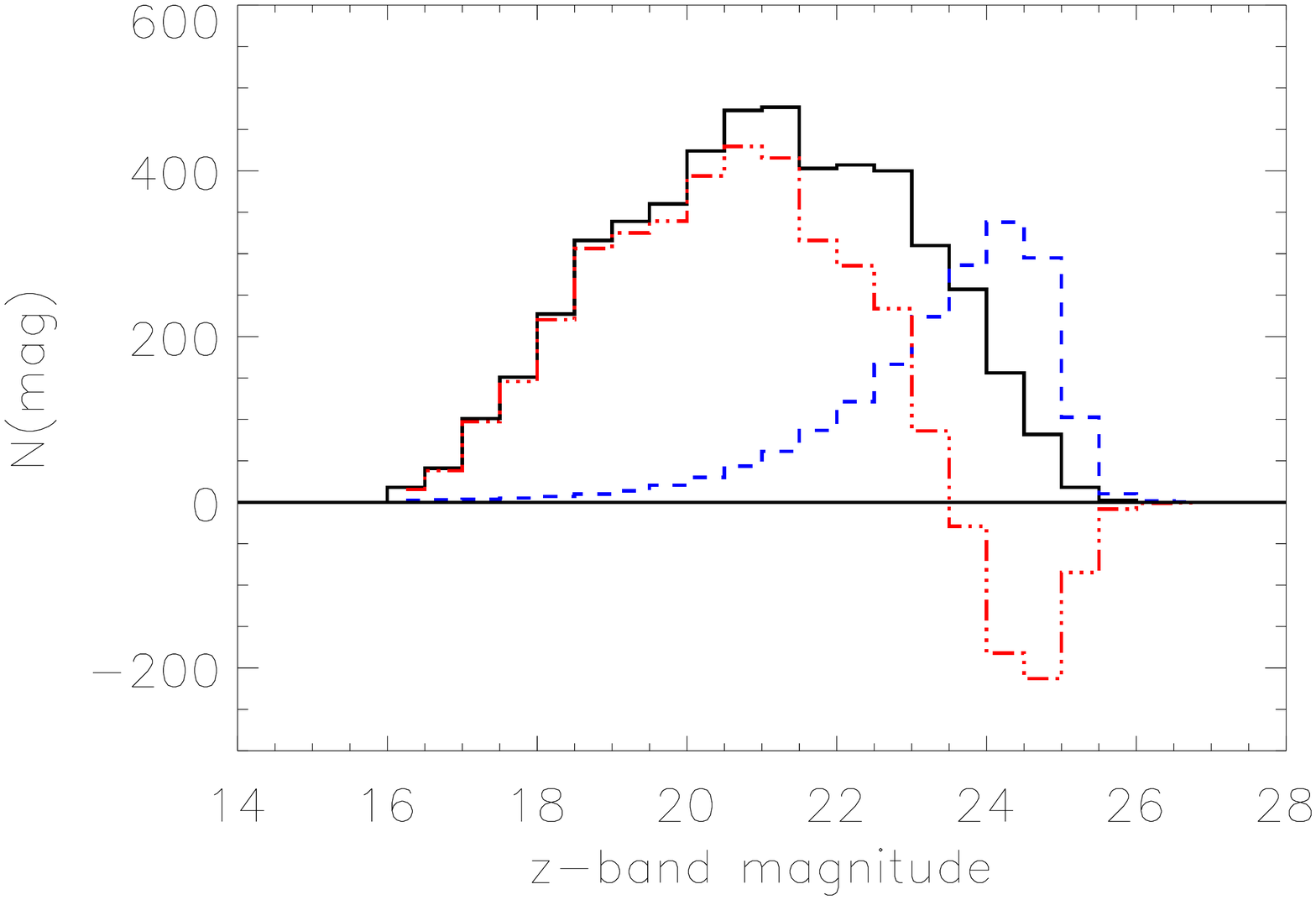} \\
        \caption{Observed magnitude distribution of all optical objects present in the $z_{DEC}$ band catalogue within a
                                radius of 3\arcsec around each radio source is shown by the black solid histogram. The expected distribution of background
                                objects in the same area  $n(m)$ is shown by the blue dashed histogram. The difference 
                                between these two distributions ($q(m)$) is shown by the red dot-dashed  histogram. At magnitudes fainter than z=23.5 
                                 the $q(m)$ histogram has negative, unphysical values. }
                                 \label{fig1}
        \end{figure}

With the procedure described in the previous section, $q(m)$ is well defined up to  magnitudes of approximately $23.0-23.5$, depending on the band. 
At fainter magnitudes, the number of objects around the radio sources turned out to be smaller than  expected from the background global counts $n(m)$.
This would result in an unphysical negative $q(m)$, which would not allow the application of this procedure at these magnitudes.
This effect can be clearly seen in the magnitude distributions of Fig.~\ref{fig1}, where for the  DECam   catalogue in the $z$ band (z$_{DEC}$), 
we report the background magnitude distribution $n(m)$ (blue dashed  histogram),  the observed magnitude distribution around each radio source
within a radius of 3 arcsec (black solid histogram), and the expected magnitude distribution of the true optical counterparts $q(m)$ (red dash-dot-dot-dot histogram)
calculated as the difference of the two previous distributions.   At magnitudes fainter than $z_{DEC}$=23.5, we have the unphysical situation that the observed magnitude  
distribution around the radio sources  is  lower than the value of the background distribution.  This effect was already noted by Brusa et al. (2007)
during the identification of the X-ray sources in the COSMOS field.

The reason for this effect is the presence of a large number of relatively bright optical counterparts close to the radio sources. 
These objects make it difficult to detect fainter background objects in the same area. 
As a consequence, at faint magnitudes, the background distribution  
calculated  around  the radio sources is underestimated when compared to the background calculated from the global field.

\begin{figure} 
        \includegraphics[width=\columnwidth]{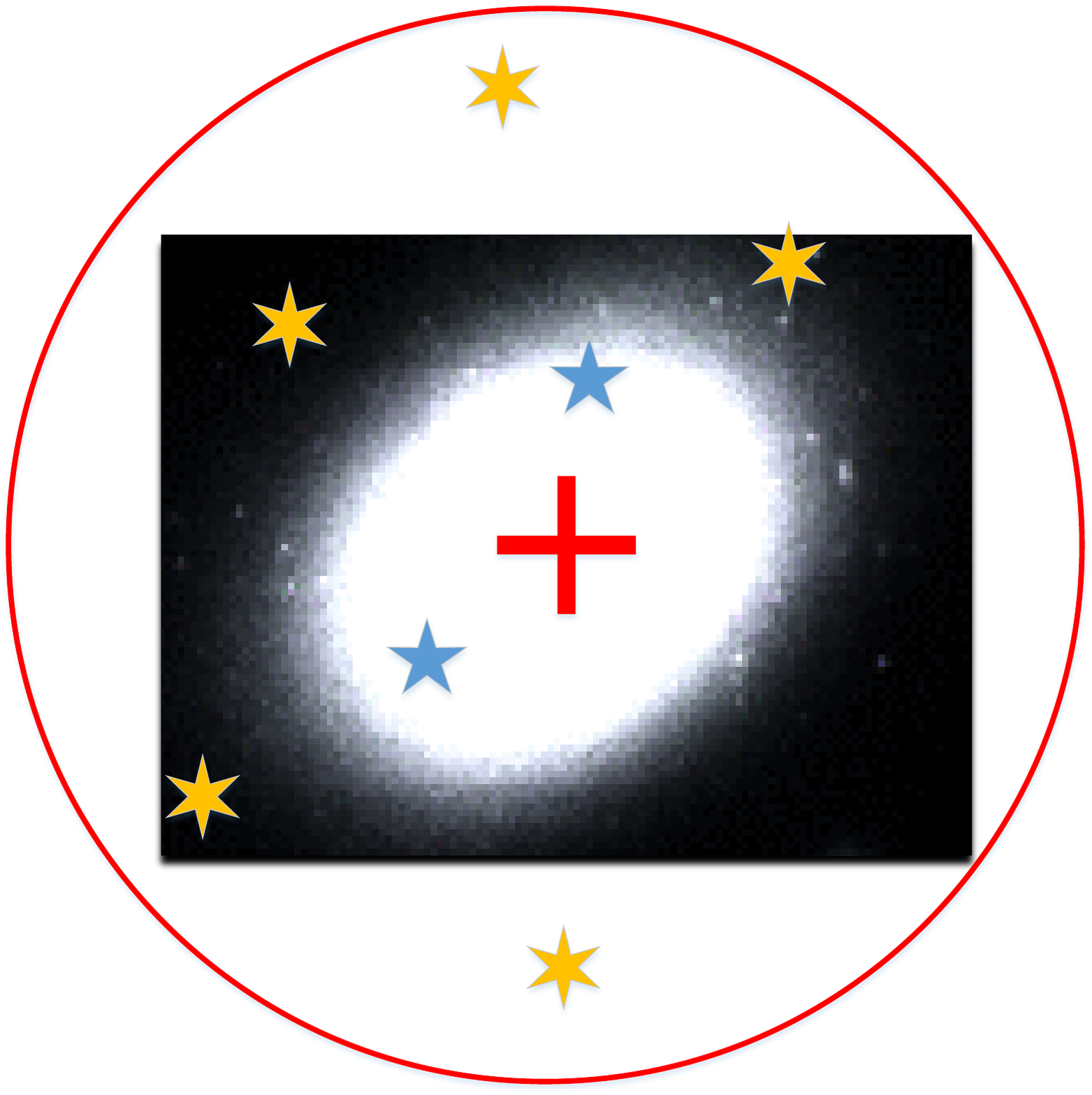} \\
        \caption{Visual representation of the effect that produces   negative values at faint magnitudes of the expected probability distribution of the 
           true optical counterparts.  The large red circle represents the area used to estimate the observed magnitude distribution 
           around the radio source, whose  position  is represented by the red cross. Background sources are represented by yellow six-pointed stars and blue five-pointed 
           stars.  Due to the presence of a large number of relatively bright optical counterparts (represented in the figure),  faint 
           background objects in the area masked by the bright sources (blue five-pointed stars) are not detected.  This 
           causes an underestimation of the observed magnitude distribution  calculated around the position of the radio sources.   }
           \label{likel_design}
           \end{figure}

This situation is visually represented in  Fig.~\ref{likel_design} , where a bright optical galaxy has been assumed as the optical counterpart 
of a radio source,  the position of which is represented by the red cross.  The large red  circle represents the area used to estimate the observed 
magnitude distribution around the radio sources (a circle of radius 3 arcsec in our case).  In this example there are eight sources that should contribute 
to the observed magnitude distribution:  the bright optical galaxy (counterpart of the radio source) and seven background sources represented 
by the seven stars in the figure (five yellow six-pointed stars plus two blue five-pointed stars).   However, while the bright galaxy and the five six-pointed sources
are  easily detected and will contribute to the observed magnitude distribution,  the two blue five-pointed  stars cannot be detected because they are in the 
area masked by the bright galaxy. This causes  an underestimation of the magnitude distribution of faint sources. 

To estimate the correct $n(m)$ to be used at faint magnitudes in the likelihood calculation, following Brusa et al. (2007) we used the technique described below. 
We have randomly extracted 5000 sources from the optical sample with the same magnitude distribution expected for the optical counterparts. 
Then we computed the background surface density around these objects. 
The $n(m)$ computed with this method is consistent with the global $n(m)$ for brighter magnitudes (approximately up to $23.0-23.5$), but it is significantly 
lower at fainter magnitudes. Therefore, for brighter magnitudes we used the global value as the input
$n(m)$ in the
likelihood procedure, while for the fainter magnitudes we used the $n(m)$ computed as explained above. 
This allowed us to identify as probable counterparts a few tens of very faint sources that would have been missed without this correction in the expected $n(m)$.

\begin{figure} 

        \includegraphics[width=\columnwidth]{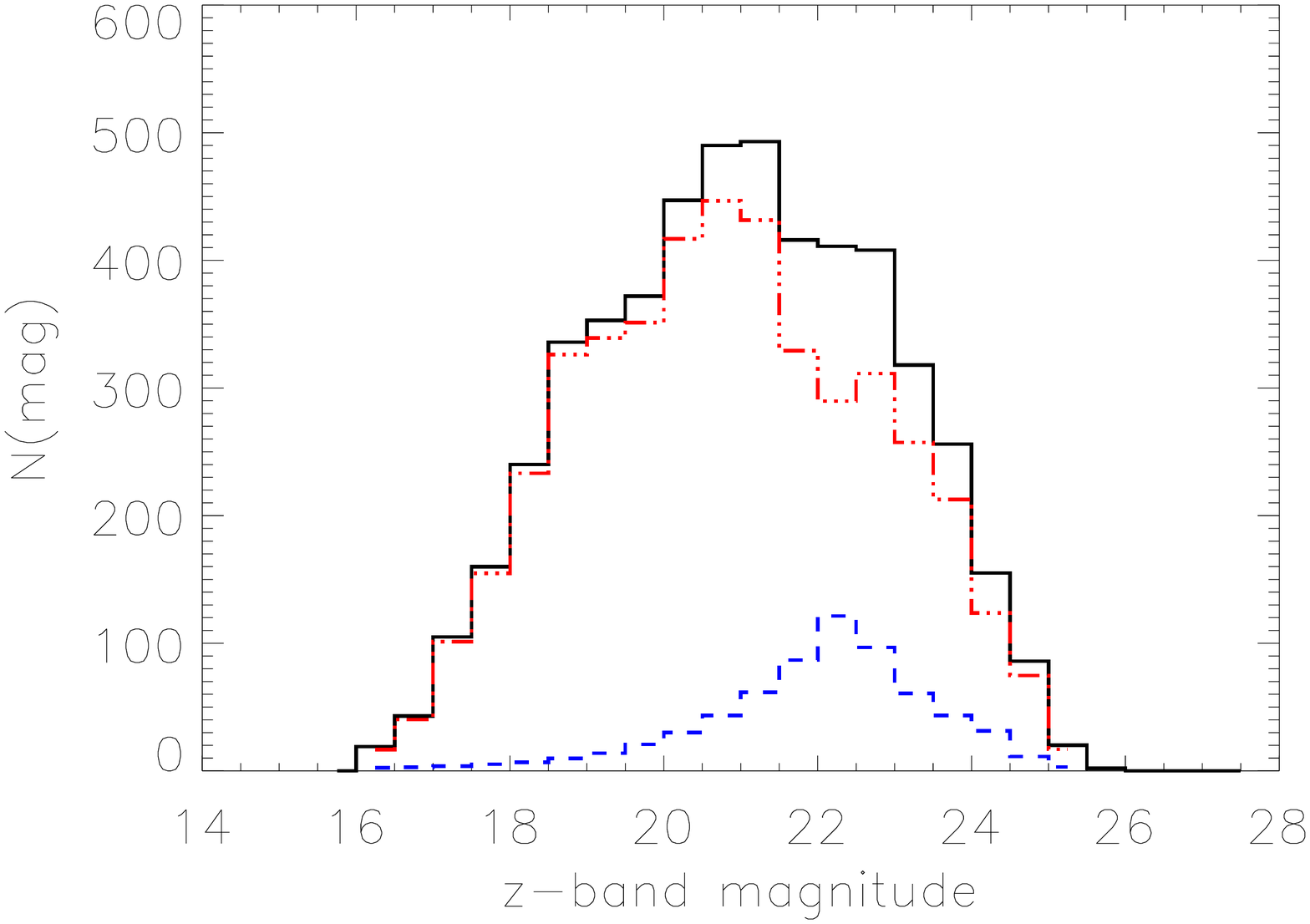} \\
        \caption{Observed magnitude distribution of all optical objects present in the z$_{DEC}$ band catalogue within a radius of 3 arcsec around each radio source is shown by the solid black histogram. The expected distribution of background objects in the same area, estimated using the procedure described above, $n(m)$ is shown by the blue dashed histogram. The difference between these two distributions, $q(m)$, is shown by the red dot-dashed  histogram}
        \label{lik_final}       
\end{figure}

In Fig.~\ref{lik_final} the new  $n(m)$ and $q(n)$ distributions calculated with the new method explained above are shown.  As shown in the figure, 
now the observed (solid line)  magnitude distribution  around the radio sources (background  and counterparts of radio sources) is, as expected, always  greater than the background distribution (dashed line), giving a  $q(m)$ distribution  without negative values.  

\subsection{Counterpart selection procedure}

The presence or absence of more than one optical candidate for the same radio source gives us additional information to that contained in $LR$. 
The reliability $Rel_{j}$ for object $j$ being
the correct identification is given by the equation (Sutherland \& Saunders, 1992)
\begin{equation}
Rel_{j} = \frac{LR_{j}}{\sum_{i}(LR)_{i}+(1-Q)},
\end{equation}
where  the sum $\sum_{i}$  is over the set of all candidates for this particular source, while $Q $ is the probability that the optical/NIR  counterpart of
the source is brighter than the magnitude limit of the optical catalogue ($Q = \int^{m_{lim}} q(m)dm$). 
The $Q$ value has been estimated by the comparison between the expected number of identifications 
derived from the integral of the $q(m)$ distribution (red dot-dashed  histogram in Figure 3) and the total number of the radio sources.  The $Q$ value has been 
calculated for each band considered  and  we find that a value of $Q$=0.8 is applicable to all  bands.   \\
Once $q(m)$, $f(r)$, and $n(m)$ were obtained, we computed the $LR$ value for all the optical sources within a distance of 3\arcsec
~from the radio position. Having determined the $LR$ for all the optical candidates, we had to choose the best threshold value
for $LR$ ($LR_{th}$) to discriminate between spurious and real identifications. As the LR threshold we adopted $LR_{th} = (1-Q)=0.2$. With this value,
according to Eq. (2) and considering that we assumed $Q$ equal to 0.8, all the optical counterparts of radio sources with only
one identification (the majority in our sample) and $LR > LR_{th}$ have a reliability greater than 0.5.  This choice also approximately maximises the sum of 
sample reliability and completeness and has been adopted also in previous paper (Ciliegi et al. 2005).


\section{Results of the likelihood ratio technique}

\begin{figure}
 \centering
      \includegraphics[width=4.2cm,clip]{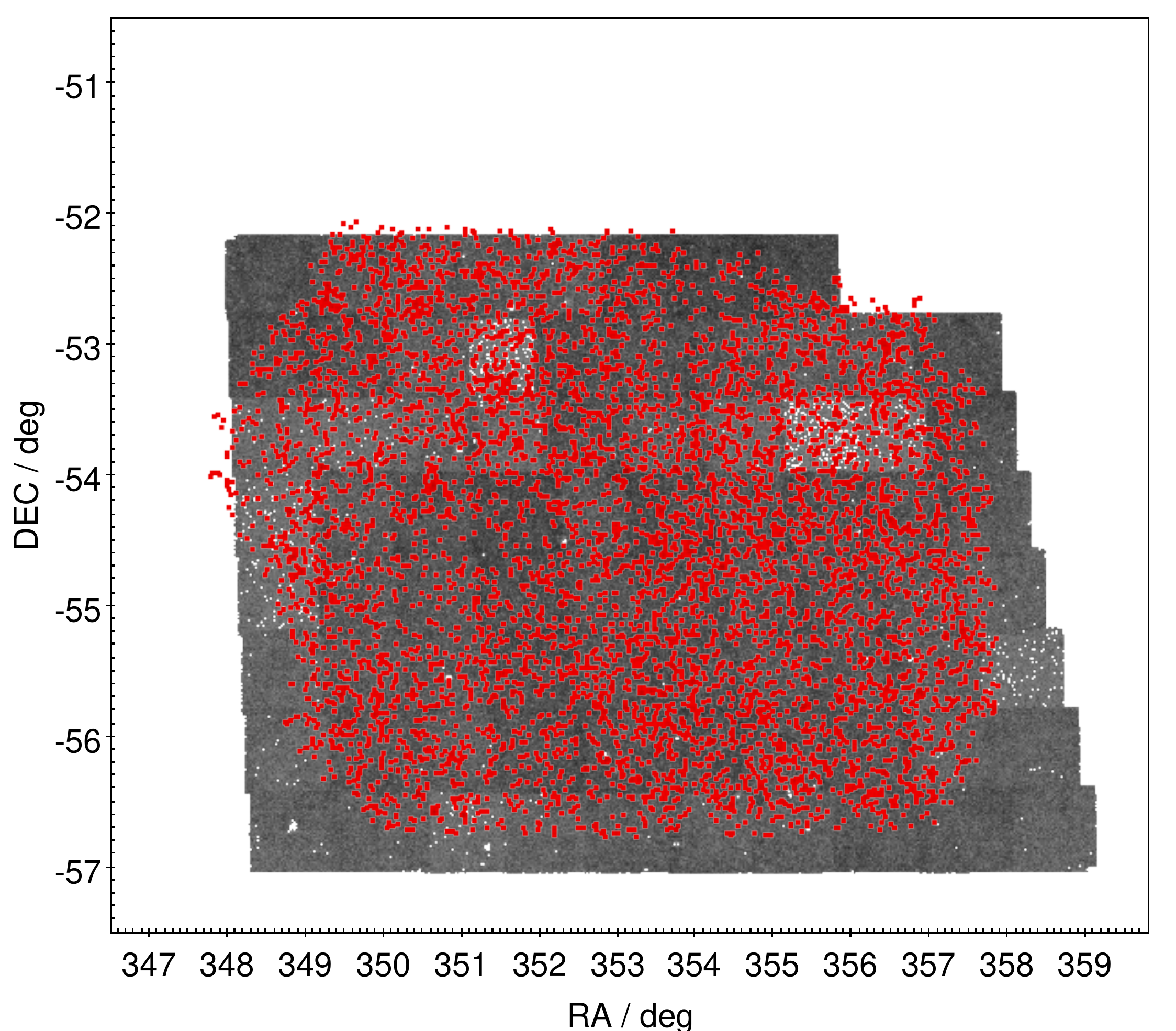}
      \includegraphics[width=4.2cm,clip]{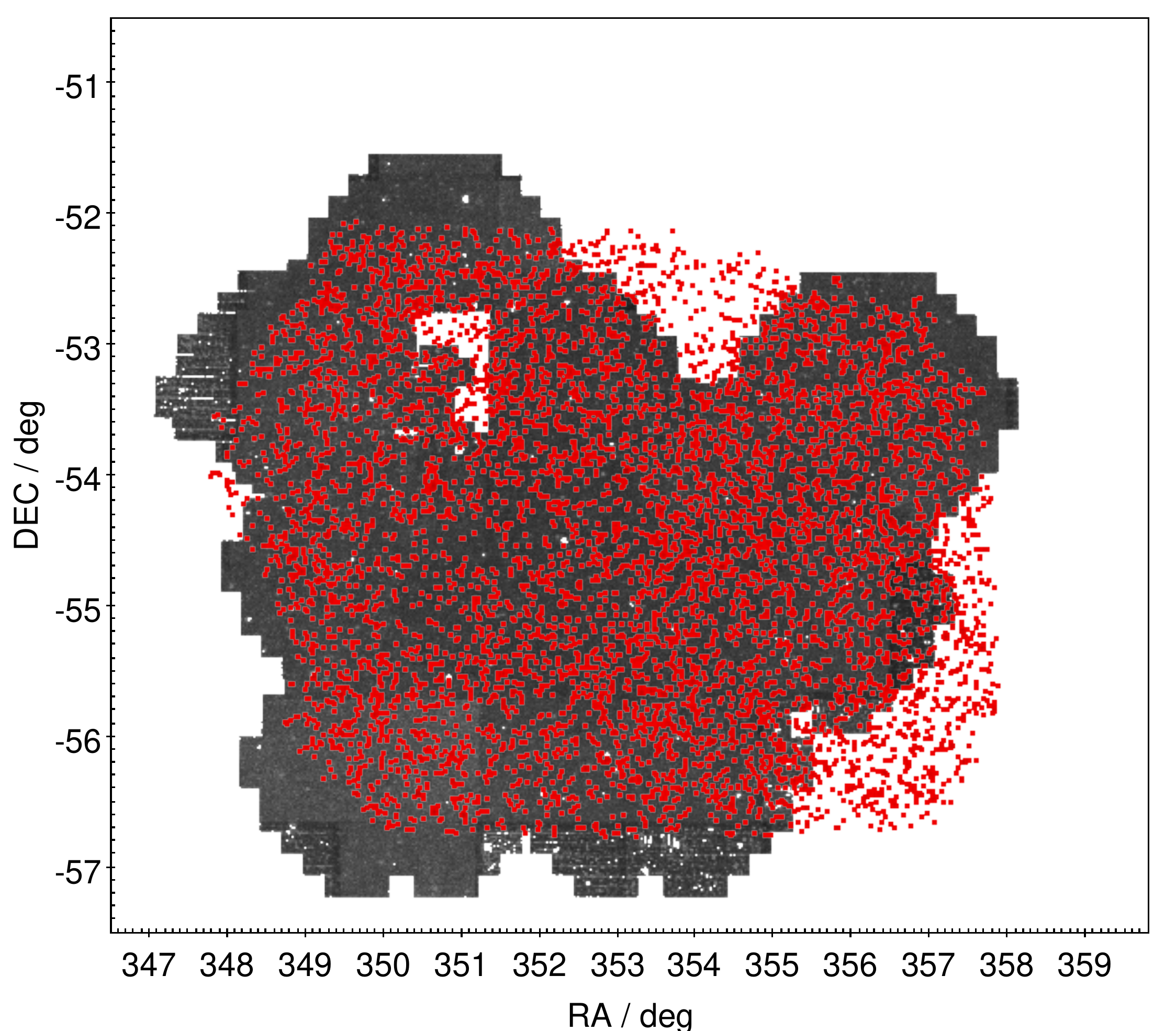}
      \includegraphics[width=4.2cm,clip]{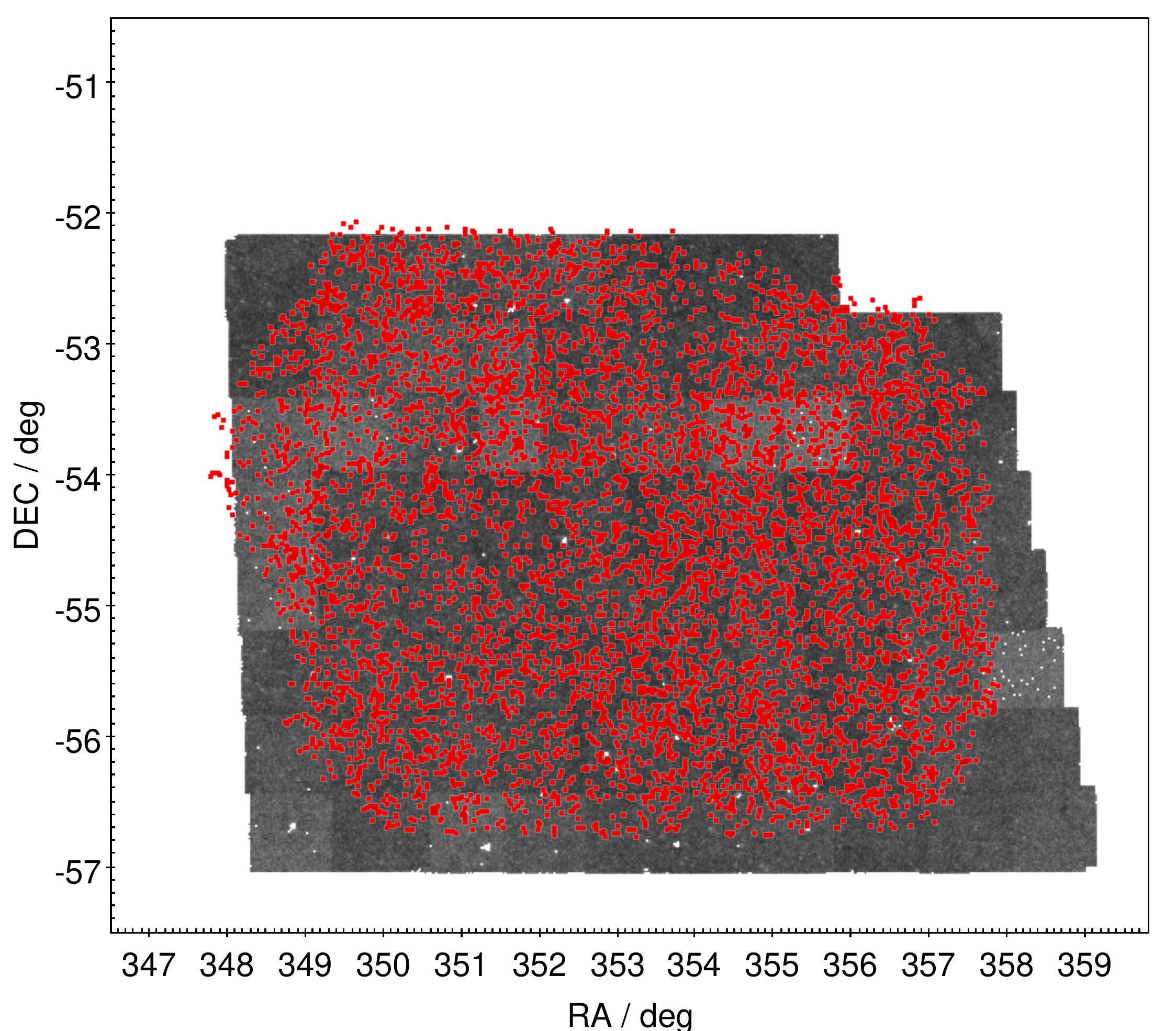}
      \includegraphics[width=4.2cm,clip]{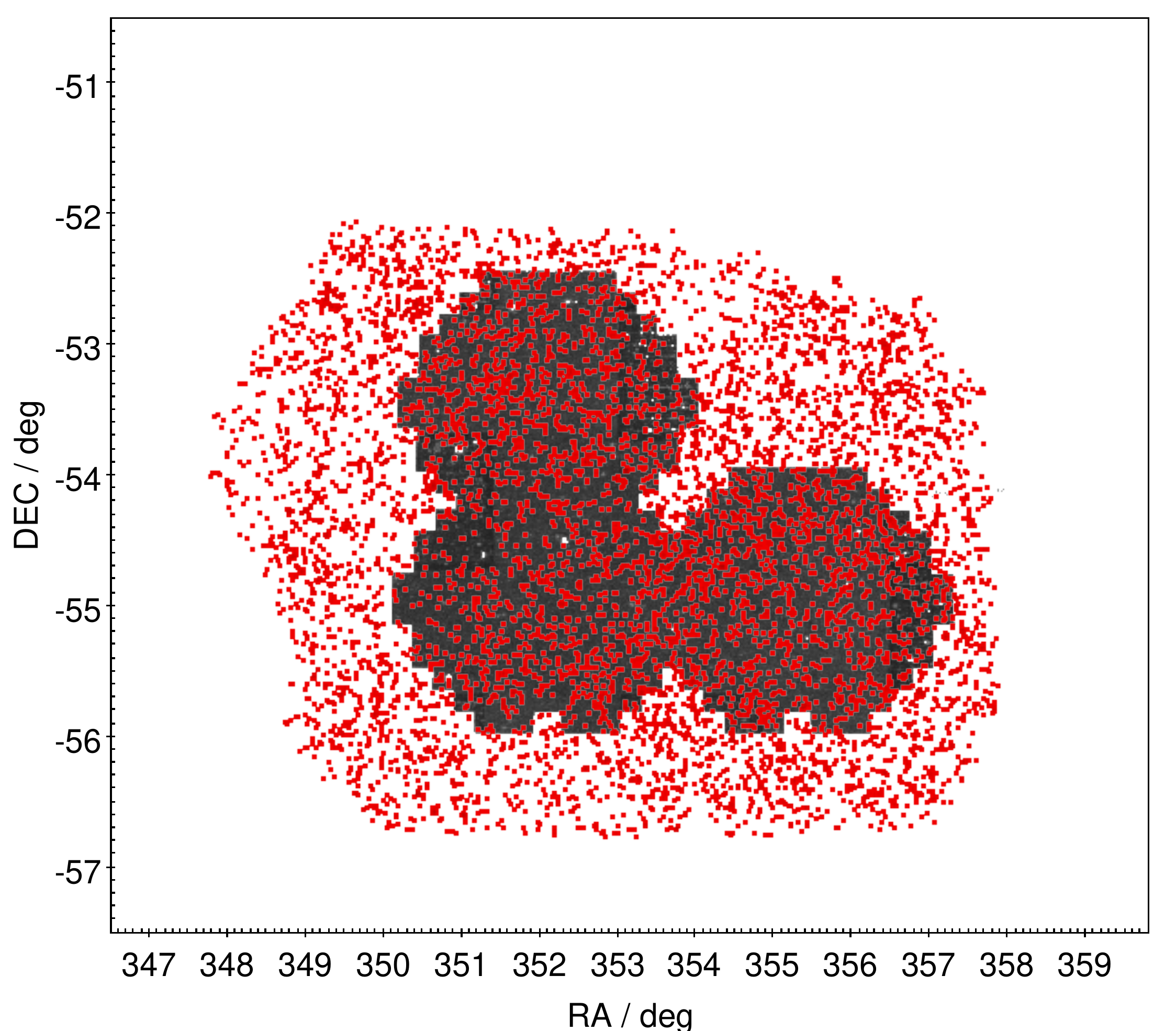}
      \includegraphics[width=4.2cm,clip]{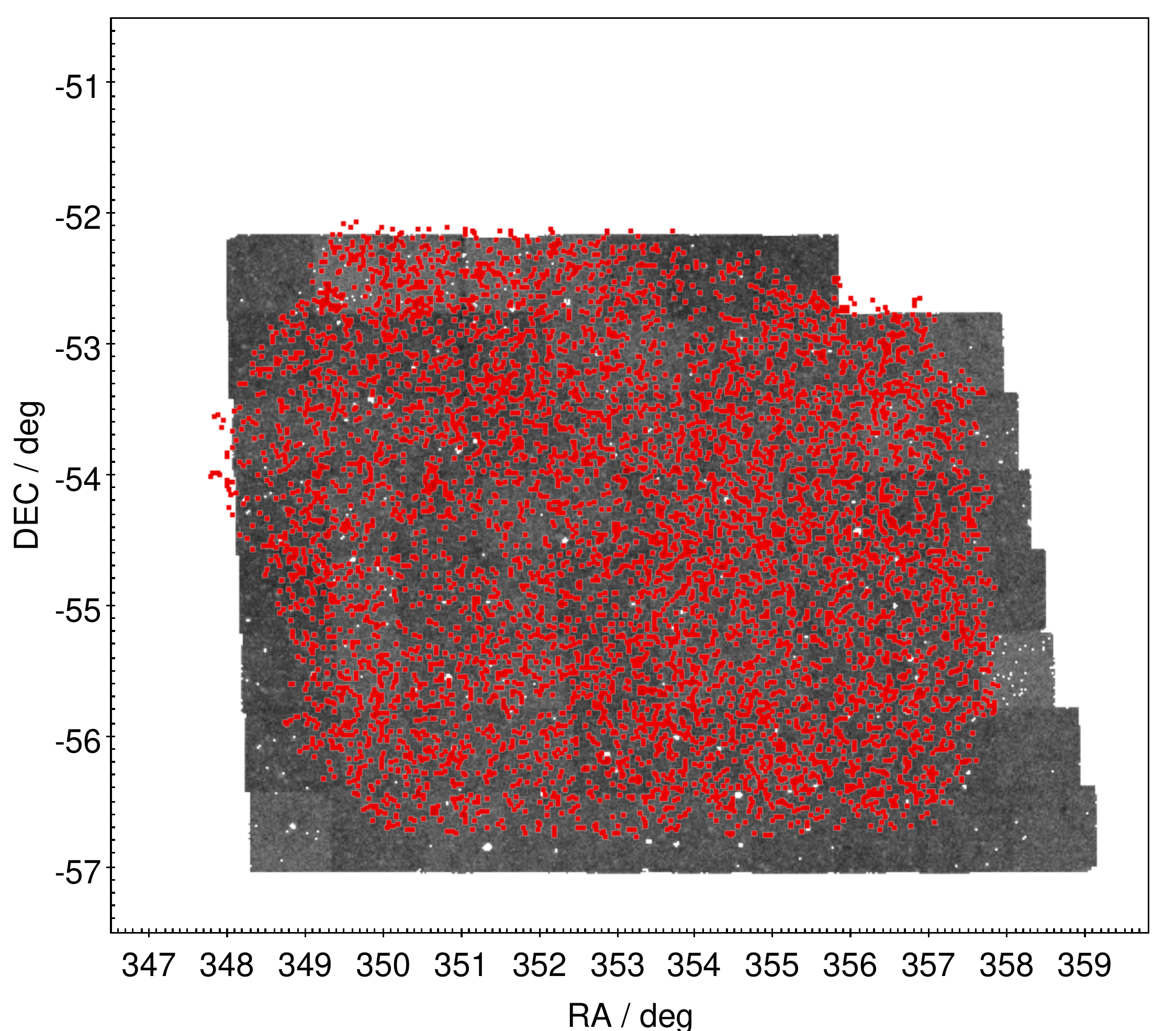}
      \includegraphics[width=4.2cm,clip]{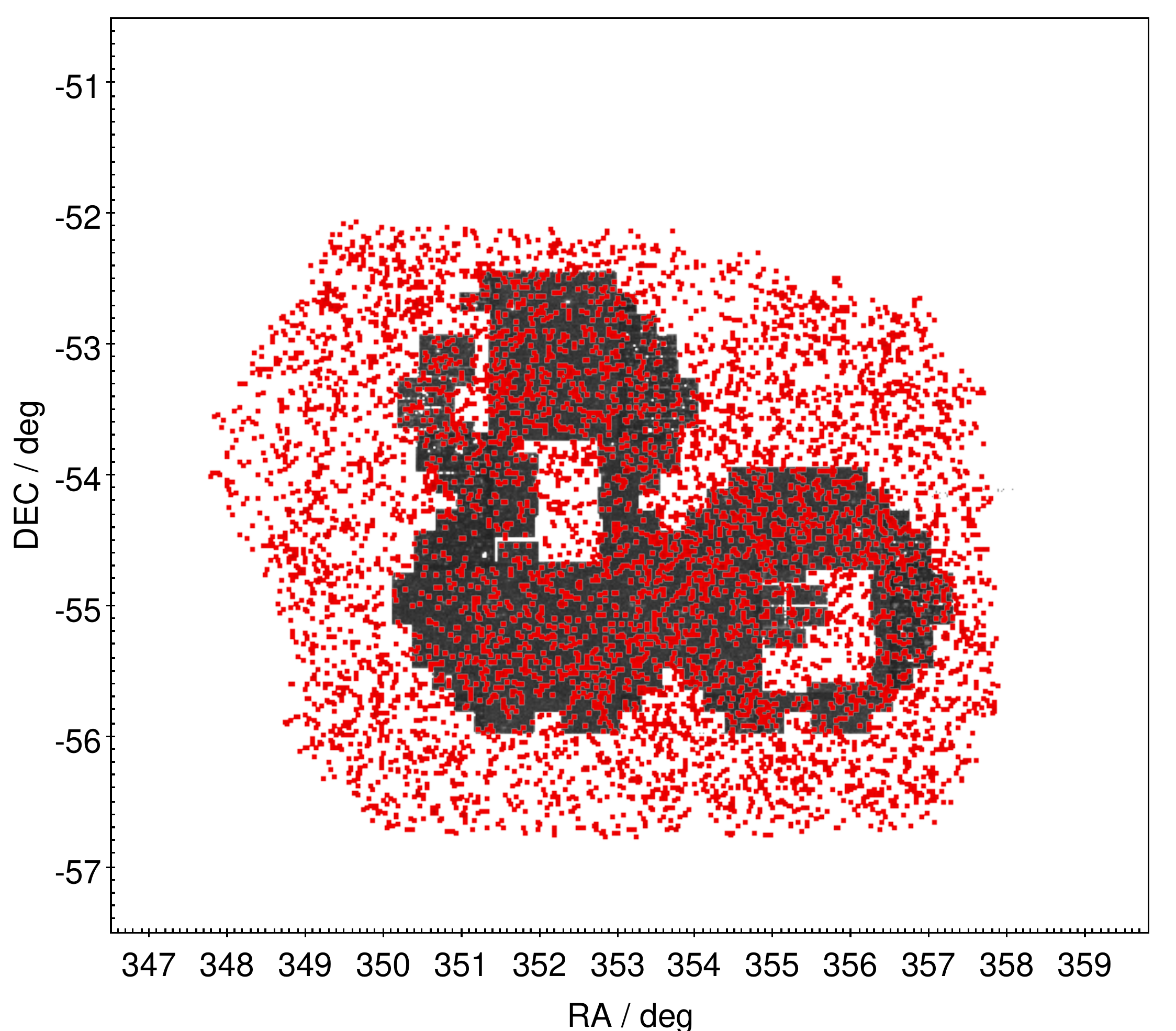}
      \includegraphics[width=4.2cm,clip]{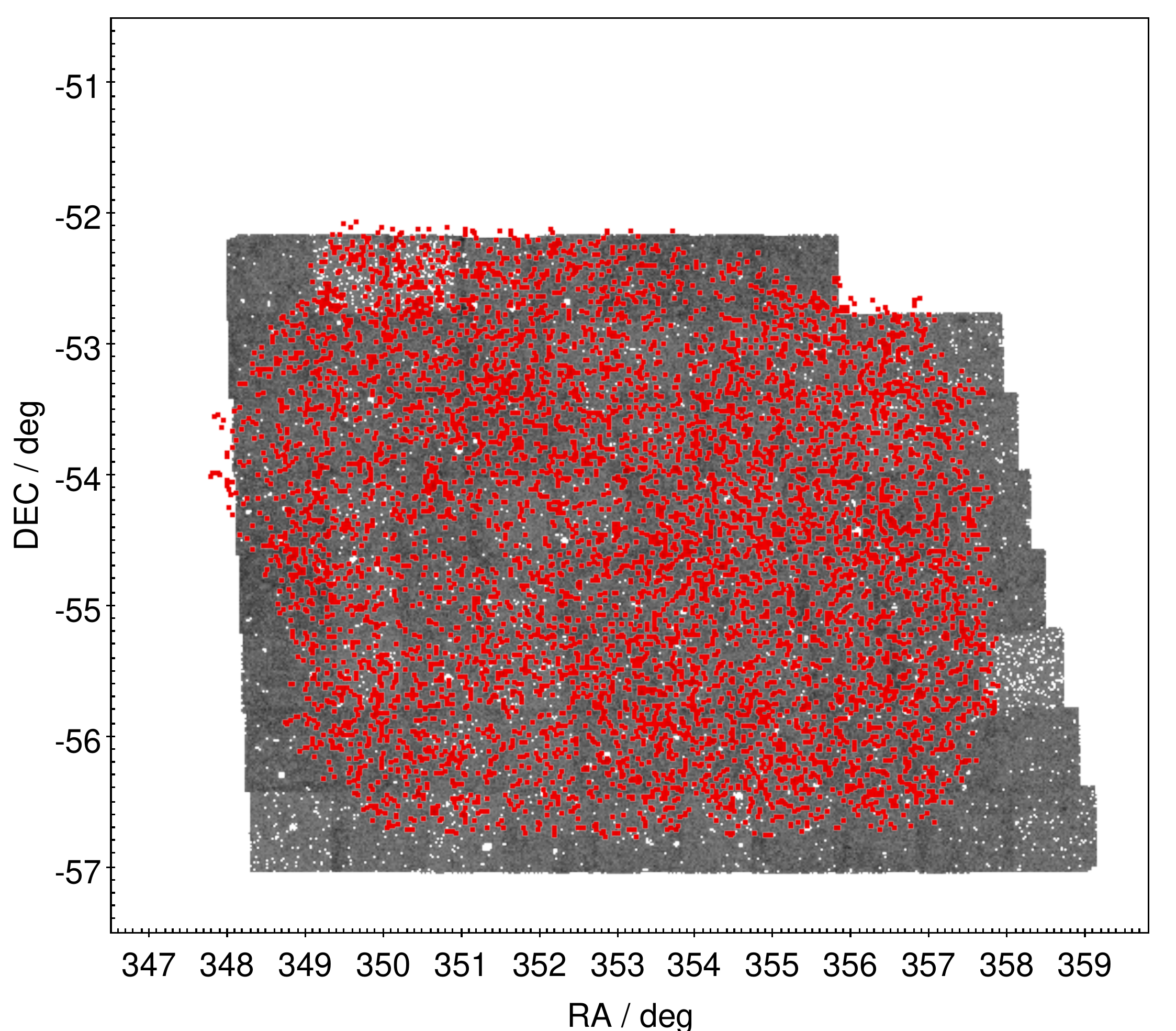}
      \includegraphics[width=4.2cm,clip]{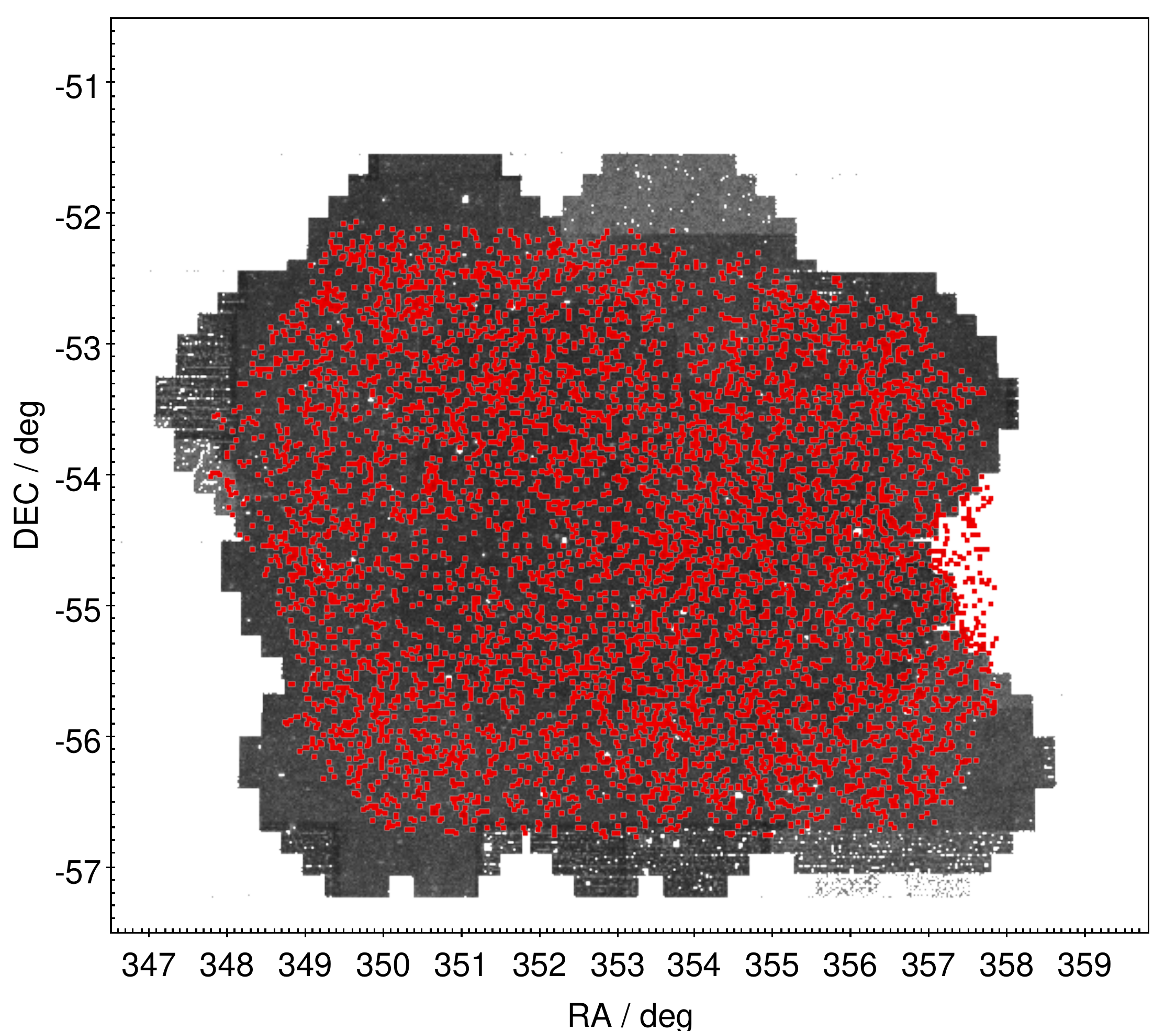}      
     
      \caption{Overlap between the position of the optical sources (black tiles) and radio sources (red points). From top left to bottom right the bands are  
         $g_{BCS}$, $g_{DEC}$, $r_{BCS}$, $r_{DEC}$, $i_{BCS}$, $i_{DEC}$, $z_{BCS}$, and $z_{DEC}$. }
         \label{ra_dec_opt}                    
                
   \end{figure}

\begin{figure}
    \includegraphics[width=4.2cm,clip]{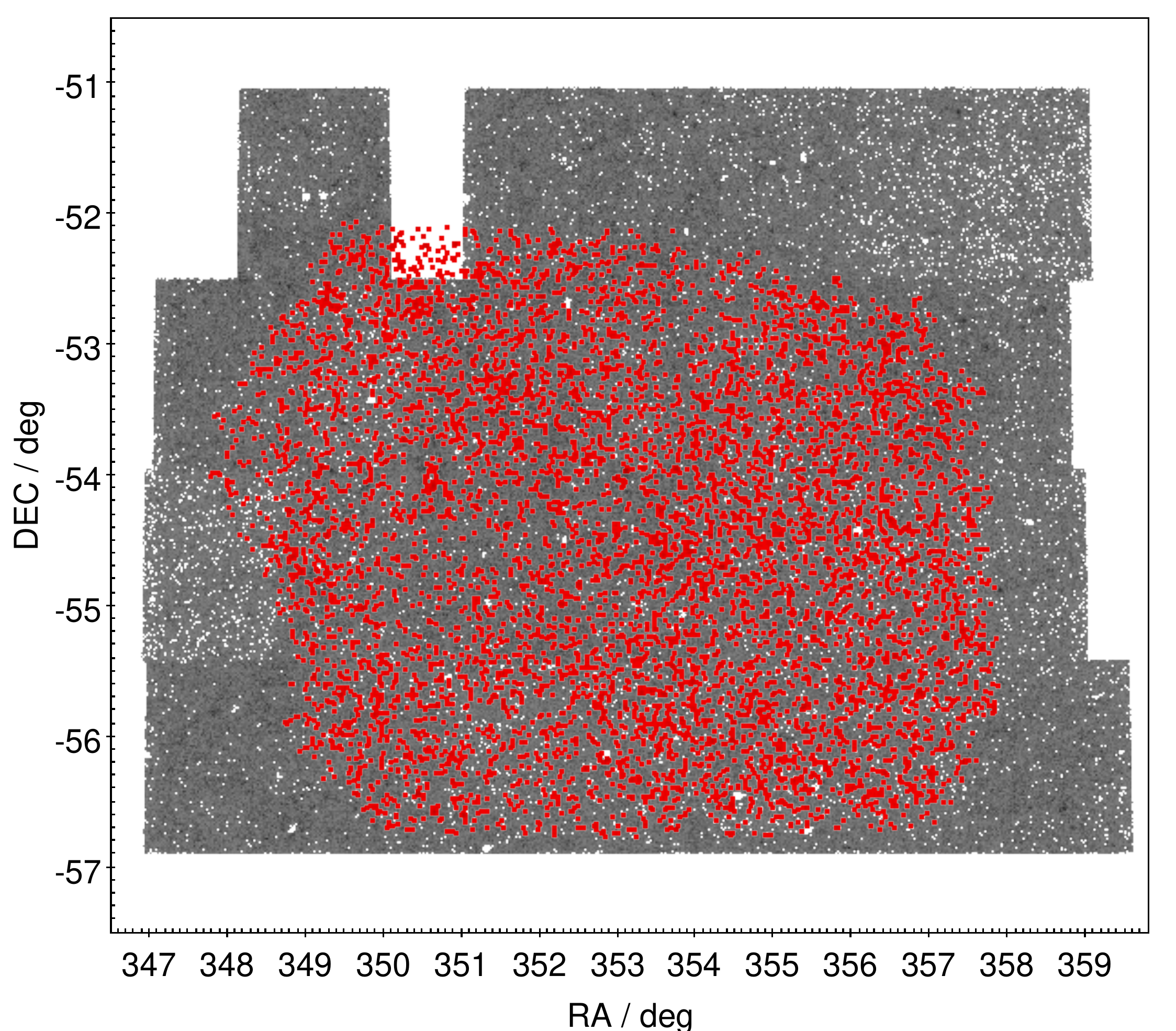}
      \includegraphics[width=4.2cm,clip]{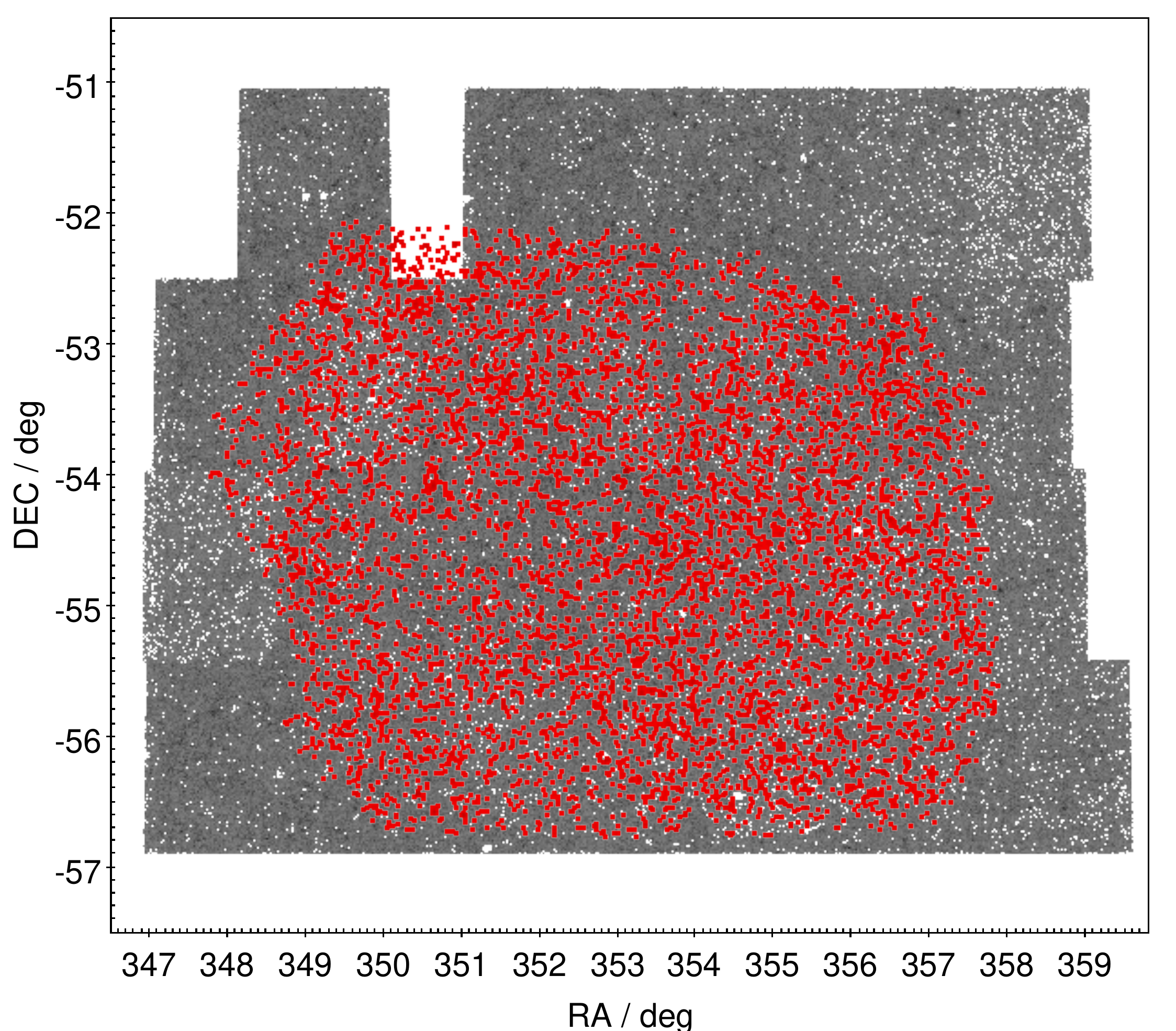}
       \includegraphics[width=4.2cm,clip]{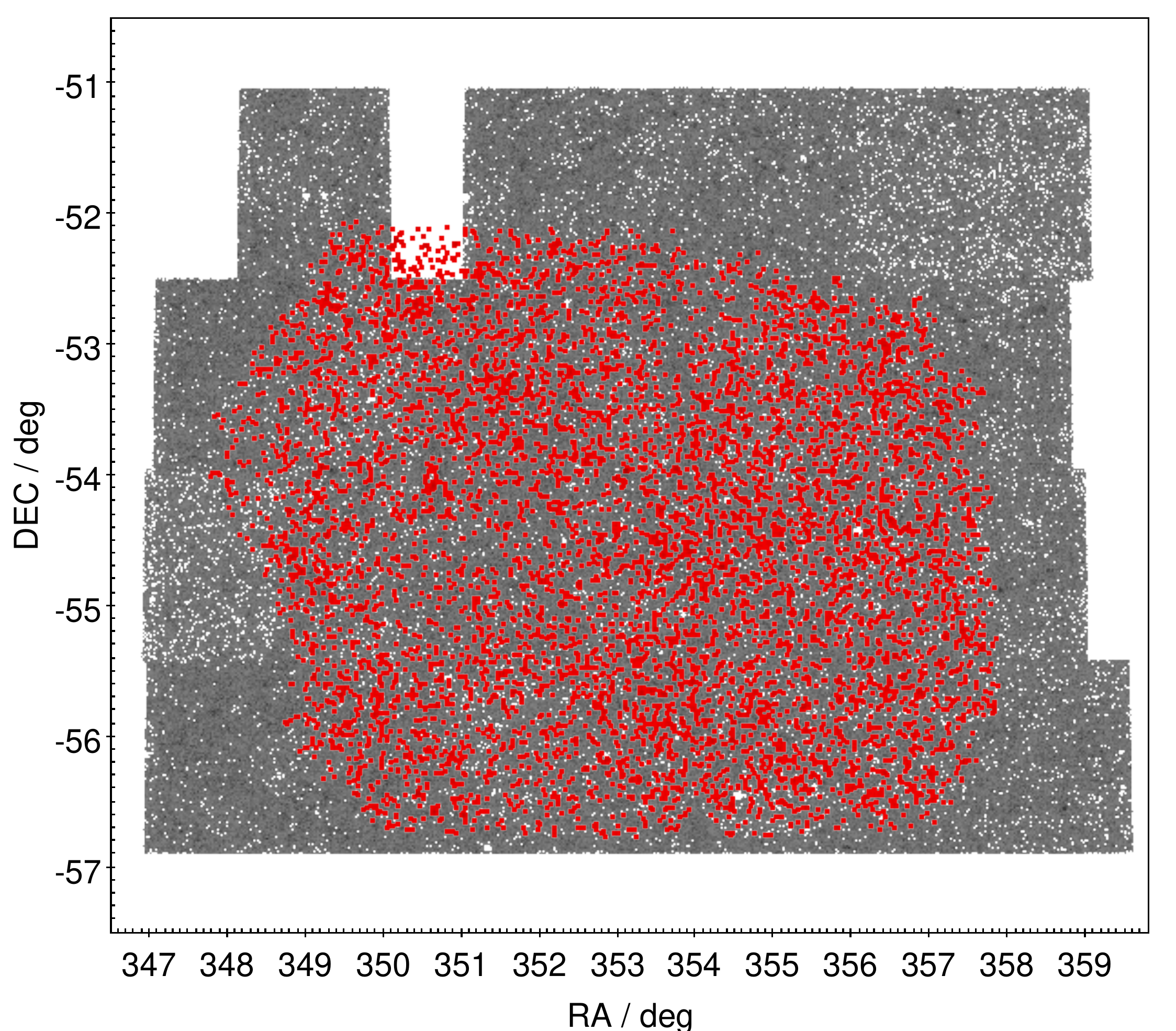}               
      
                \caption{Overlap between the position of the NIR sources (black tiles) and radio sources (red points). Top left J band, top right H band, bottom panel K band.}
         \label{ra_dec_nir}
\end{figure}

The optical/NIR photometric data cover each a slightly different area of the sky.   In Figs.~\ref{ra_dec_opt} and ~\ref{ra_dec_nir} we show the overlap of the optical and radio positions for all the bands considered. 
As is evident in the figures,  while the NIR data in the $J$, $H$, and $K$ bands (Fig.~\ref{ra_dec_nir}) cover the radio sample in a uniform way with only few radio sources left out,  the optical bands 
do not cover the radio sample in a homogeneous way.     All the data from the BCS survey ($g_{BCS}$, $r_{BCS}$, $i_{BCS}$, and $z_{BCS}$ ) cover almost the whole radio sample (Fig.~\ref{ra_dec_opt})
 at a shallower  magnitude limit (see Section 2.2), while the DEC data in the $g$, $r$, and $i$ bands  cover only partially  the radio sample (Fig.~\ref{ra_dec_opt}).  Finally the $z_{DEC}$ 
data offer the best data set in terms of covered area and magnitude limit.    In order to 
  maximise the number of identifications,  we applied the LR technique to all the optical and NIR available data, excluding only the $z_{BCS}$ data since the $z_{DEC}$  data cover the same area  at deeper magnitude limit. 
In summary, the LR technique has been applied to seven optical ($g_{BCS}$, $g_{DEC}$, $r_{BCS}$, $r_{DEC}$, $i_{BCS}$, $i_{DEC}$, $z_{DEC}$) and three NIR ($J$, $H$, $K$) bands. 

%

Considering all the available data, 6176 radio sources (from a total  of 6287 radio sources) 
are inside the area covered  by at least one optical or NIR band, i.e. only  for 111 radio sources is it   impossible to obtain the information on the optical/NIR counterpart because we have no optical/NIR data for them. 

\subsection{Identified radio sources} 

The results of the identification process  are  summarised  in Table ~\ref{table:1}.   
For each band, we report the magnitude limit, the number of radio sources  within the area covered by the corresponding optical or 
NIR band, the number of  reliable identifications obtained using the LR technique, and the relative fraction of identification.
As expected, the fraction of  identification is function of the depths of the optical/NIR data available, varying from 48.3\% for the shallower data ($g_{BCS}$) to 
 72.2\% for the the deepest data ($z_{DEC}$).  These fractions of
identification  are consistent with previous radio-optical association studies at similar optical magnitude depths.  Ciliegi et al. (2005)  identified 74\% of the radio sources in the VIMOS-VLA Deep Survey (VVDS) field using optical data with a magnitude limit  of $I_{AB}\sim$25.0. 
However, these fractions of identification are 
relatively low in comparison to similar work conducted in other radio fields using deeper optical and NIR data. 
Recently, for example, Smol{\v c}i{\'c} et al. (2017a)  have found counterparts for 93\% of the 3GHz radio sources in the COSMOS field using photometric catalogues in the $Y, J, H, K_{s}$ bands down to a 3$\sigma$ (3$^{\prime\prime}$ diameter aperture) of 25.3, 25.2, 24.9, and 24.5 AB magnitude.

 The ten different catalogues have been combined in a single master catalogue of optical/NIR counterparts using the following procedure: 
 
 \begin{enumerate}

\item When a radio source is associated with the same counterpart in all the catalogues,  we assign this counterpart to the radio source in the final master catalogue;   
\item When a radio source is associated with different counterparts in different optical/NIR catalogues, we may have two different cases: 
\subitem a) all the  counterparts are detected in at least one common optical/NIR band (e.g. all the counterparts  are detected in the $z_{DEC}$ catalogue).  There are 73 radio sources in this case; 
\subitem b)  none of the  counterparts has a detection in a common band (e.g. one  counterpart is  detected in the $z_{DEC}$ band and  the other counterpart is detected in the $g_{DEC}$ band but not in the $z_{DEC}$  band). There are 373 radio sources in this case. 
In the first case we selected the counterpart of the radio sources on the basis of  reliability values in the common band, while in the second  case we selected the  nearest counterpart to the radio source because it is  impossible to compare reliability values from different bands. 
\end{enumerate} 

This  procedure led to the identification of 
optical/NIR counterparts for 4770 different radio sources, i.e. for  77\% (4770/6176) of the entire radio sample.    
Although the final master catalogue is  inhomogeneous in terms of  depth and area covered (see Fig.~\ref{ra_dec_opt} and Fig.~\ref{ra_dec_nir}), it is an important tool for a detailed 
photometric analysis  of the counterparts of the radio sources (see XXL Paper XXXI).

Among the 4770 optical/NIR counterparts of the XXL radio sources,  there are 414  
sources with an X-ray counterpart that had flux measurements in at least one $XMM$ band (see XXL Paper XXXI, and Fotopoulou et al. in prep)

\begin{table*}
\caption{Results of the identification process using the likelihood ratio technique in seven different bands} 
\label{table:1} 
\begin{tabular}{l l l l l } 
\hline\hline
Band & Magnitude        & No. of radio sources & No. of reliable & Fraction of \\
          & Limit                 & within area covered by & identifications            & identification \\
          & (AB mag)$^1$ & optical/NIR data            &                                &                       \\

\hline 
 $g_{BCS}$   &    24.14   &   6041  &  2919 &  48.3 \%   \\
 $g_{DEC}$   &    25.73   &   5492 & 3414 &  62.1 \% \\
 $r_{BCS}$    &    24.06   &   6183  & 3742 & 60.5 \% \\
 $r_{DEC}$    &    25.78  &    2732  & 1902 & 69.6 \%  \\
 $i_{BCS}$    &     23.23 &    6139  &  3837 &  63.1 \% \\
 $i_{DEC}$   &     25.60  &    2414  &  1724 & 71.4 \% \\
  $z_{DEC}$  &     24.87  &     6132 &  4433 &  72.2 \% \\
 $J$              &     21.10  &    5762  &   3548 & 61.5 \% \\
 $H$             &     20.77  &    5781 &  3714 &  64.2 \% \\
 $K$             &     20.34  &      5719 &  3823 &  68.8 \% \\
\\
\hline 

\end{tabular}

$^1$ See section 2.2 for more details

\end{table*}

Table 2 shows representative entries for the optical/NIR  and X-ray counterparts of the XXL-S radio sources.   For each identified radio source we report
the IAU name (column I)  and the ID number (column II)  as in the radio catalogue published in XXL Paper XVIII,  the radio peak flux density $S_p$ corrected for the bandwidth smearing (column III),  
the integrated flux density $S_{int}$ (column IV), and  the separation in arcsec
between the radio position and the optical/NIR position of the associated counterpart (column V).  Multicomponent radio sources have $S_p$=$-$99 (see XXL Paper XVIII), while unidentified 
radio sources have a separation value (column V) equal to $-$1. From column VI to column XV we report the AB magnitude in ten 
different bands.   Values of 99.99  are for radio sources outside the area covered by the optical/NIR data.  
The 111 radio sources without optical/NIR data available (see above) all have  magnitude values equal to 99.99. Negative values represent magnitude limits (i.e. sources not detected but observed in that band). 
The magnitude value after the  minus sign gives the magnitude limit in that band (see section 2.2 and Table 1). 
Finally,   in column XVI we report the X-ray name, while in columns XVII and XVIII  we report the  X-ray flux in the 0.5--2.0  keV and 2.0--10 keV  bands.     The full catalogue is 
available as  a queryable database  table  via the XXL  Master Catalogue browser.\footnote{http://cosmosdb.iasf-milano.inaf.it/XXL}  Copies will also be deposited 
at the Centre de Donn\'ees astronomiques de Strasburg (CDS). \footnote{http://cdsweb.u-strasbg.fr} 

\begin{table*}

\caption{Example entries for the catalogue of the optical and NIR counterparts of the radio sources detected in the XXL-S field. Full table available at   the XXL  Master Catalogue browser (http://cosmosdb.iasf-milano.inaf.it/XXL) 
and at the Centre de Donn\'ees astronomiques de Strasburg (CDS) (http://cdsweb.u-strasbg.fr).  } 
\label{table:2} 
\begin{tabular}{l l r r  c r r r     } 
\hline\hline
I & II & III & IV & V & VI & VII & VIII \\
\hline
IAU NAME &  RADIO ID &  $S_p$ & $S_{int}$ &  Distance &  $g_{BCS}$ & $r_{BCS}$ & $i_{BCS}$       \\
              &       & (mJy/b) & (mJy)   & ($\prime\prime$) &  mag$^{3}$ &  mag  & mag \\
\hline 

2XXL-ATCA J232805.5$-$554110 &  90$\_$362                 &   $-$99.00   &  61.4915       &  0.763      &    24.25      &   23.48        &  22.44          \\    
2XXL-ATCA J232624.7$-$524209 & 131$\_$300                &    $-$99.00  &  59.1999       &  0.495     &   $-$24.14   &  $-$24.06   &  $-$23.23     \\ 
2XXL-ATCA J234323.7$-$560342 & 223$\_$265                &    $-$99.00  &  44.5605       &  0.494     &    22.98       &  21.76         &   21.17         \\   
2XXL-ATCA J232904.7$-$563453 & 430$\_$528$\_$2116 &    $-$99.00  &  13.7939       &  0.537     &     22.79       &  22.09        &   21.52         \\ 
2XXL-ATCA J234705.2$-$534138 & 638$\_$2971              &   $-$99.00   &  41.4072       & 0.453      &   $-$24.14    & $-$24.06   &  $-$23.23    \\    
2XXL-ATCA J233915.3$-$535746 & 844$\_$1800              &   $-$99.00   &    8.1056       & 0.468      &   $-$24.14    & 23.39         &   20.87         \\ 
2XXL-ATCA J233913.2$-$552350 &17                                &  141.99       &145.5976       & 0.193      &     16.42       & 17.33          & 17.82          \\ 
2XXL-ATCA J232248.2$-$554535 & 66                               &   19.34        &  19.9150       &  $-1.00 $ & $-$24.14      & $-$24.06    &  $-$23.23  \\
2XXL-ATCA J231656.0$-$544202 & 69                               &   25.67        & 24.9976        & 0.407      &    23.59         &  23.32        &  99.99 \\
2XXL-ATCA J233938.6$-$543908 & 898                             & 1.61            &   2.1694        & 0.732      &    17.38         & 17.31         & 17.14           \\ 
2XXL-ATCA J233607.9$-$541825 & 4710                           &  0.25           &   0.278 1       &  0.144     &     17.99        & 17.21         & 16.84            \\
 
\end{tabular}
 
 \vspace{1cm}

\begin{tabular}{ r r r r r r r c r r  } 
  
 \hline\hline
 
 IX & X & XI & XII & XIII & XIV & XV & XVI & XVII & XVIII \\
 
 \hline 
 
 $g_{DEC}$ & $r_{DEC}$ &  $i_{DEC}$ &  $z_{DEC}$ &   $J$ & $H$ & $K$ & X-Ray Name &  Bflux $^{1}$ & CDflux$^{2}$  \\ 
mag & mag &mag &mag &mag &mag &mag  &  &   \\ 

\hline 
   
24.35       &  23.72     &   22.37      &  21.61       &    20.79     &   20.36      &  19.90     & 3XLSS J232804.1-554111   &    2.87    & $-$1  \\    
24.44       &  23.54     &   22.54      &  21.69       &    20.69     &   20.14      &  19.74     &  -                                             & $-1$      & $-1$ \\ 
99.99       &  99.99     &    99.99     &  20.83       &    20.26     &   20.21      &  19.53     &  -                                             & $-1$      & $-1$ \\  
23.12       &  99.99     &   99.99      &  21.51       &     20.66    &   20.50      &  19.93     &  -                                             & $-1$      & $-1$ \\
$-$25.73  &  99.99     &   99.99      &  $-$24.87  &    18.86     &   $-$20.77 &  18.23     &   -                                             & $-1$      & $-1$ \\   
24.62       &  22.69      &  21.66      &   20.86      &     20.51    &  19.81       &  19.66     &   -                                             & $-1$      & $-1$ \\   
$-$23.73  &  $-$25.78 &  $-$25.60 &  $-$24.87 &     15.61     &   15.42     &  15.83      &  3XLSS 233913.0-552351      &  301.9 & 471.3 \\
$-$25.73  &  99.99     &  99.99       &  $-$24.87  &  $-$21.10  &  $-$20.77 & $-$20.34 & -                                  &  $-1$       & $-1$ \\  
23.65       &  99.99     &  99.99       & $-$24.87   &  $-$21.10  & 99.99        & $-$20.34 &  -                                              & $-1$      &  $-1$ \\
17.26       &  17.13     & 17.26        &  17.07       &    16.98      &  16.86      &   16.87     &   3XLSS 233938.7-543908     &    5.5    & 18.6\\    
 18.03      &  16.94     &  17.01       & 16.53        &    16.10      &     15.78   &  15.83     &   3XLSS 233607.9-541826       &   24.5   & 107.4 \\

\hline 

\end{tabular}

$^1$ Bflux :   ~0.5-2.0 keV Flux $\times$ 10$^{-15}$ erg s$^{-1}$ cm$^{-2}$  \\
$^2$ CDflux : 2-10 keV Flux $\times$ 10$^{-15}$ erg s$^{-1}$ cm$^{-2}$  \\
$^3$ All quoted magnitudes are given in the AB system. \\

\end{table*}

\begin{figure}

                \includegraphics[width=\columnwidth]{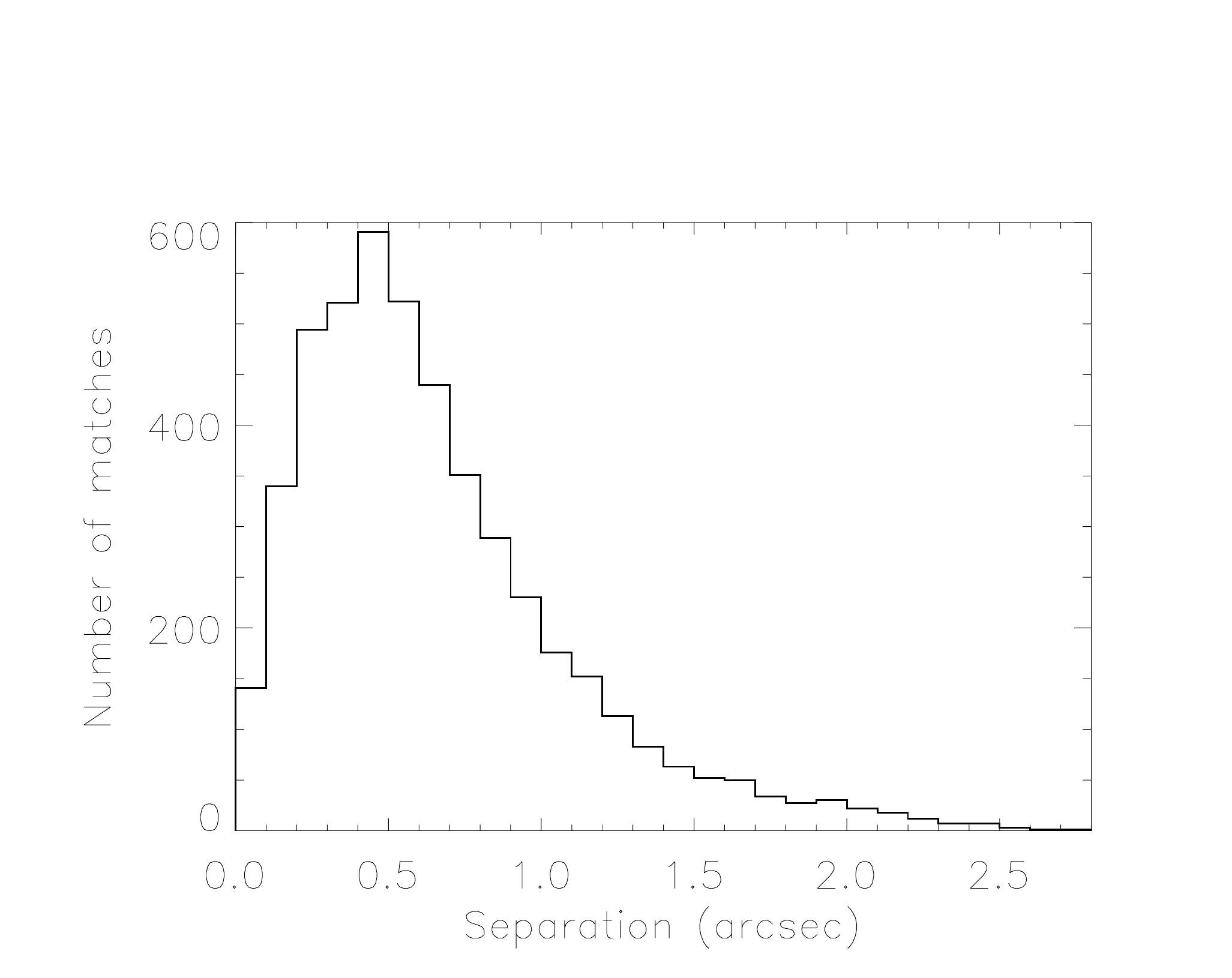} \\
                    \includegraphics[width=8.5cm,clip]{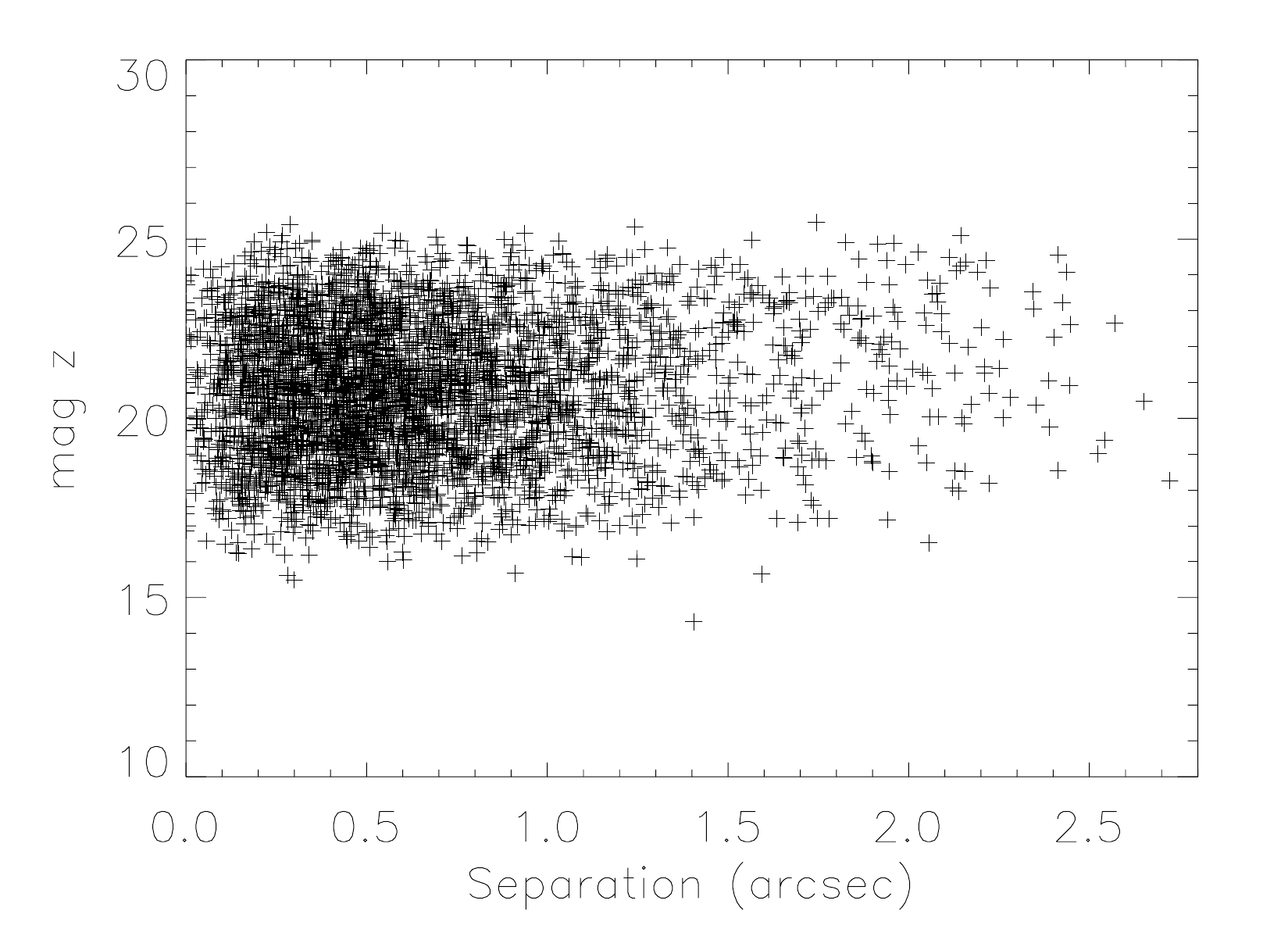} \\
                    
        \caption{$Top$: Distribution of the separation  between the radio source in the XXL-S field and their optical/NIR counterparts.  $Bottom$:  mag $z$ from the DEC survey vs  separation
         between the radio source  and their optical/NIR counterparts}  
        \label{separation}      
\end{figure}

In Fig.~\ref{separation} we report the histogram of the separation value (in arcsec) between the radio and the optical/NIR  position (column 5 in Table ~\ref{table:2}) and the mag $z$ from the DEC survey 
versus radio-optical/NIR separation.  As shown in the  figure, 
the majority of the optical and NIR sources associated with the radio sources are within a distance of 1.0 arcsec (3920 over a total of  4770 association, i.e.$\sim$82$\%$), with only 262 radio sources 
($\sim$5\%) associated with a counterpart at a distance greater than 1.5 arcsec and there is no correlation between the magnitude of the counterparts and the radio-optical/NIR separation.


\begin{center} 
\begin{figure} 
        \includegraphics[width=\columnwidth]{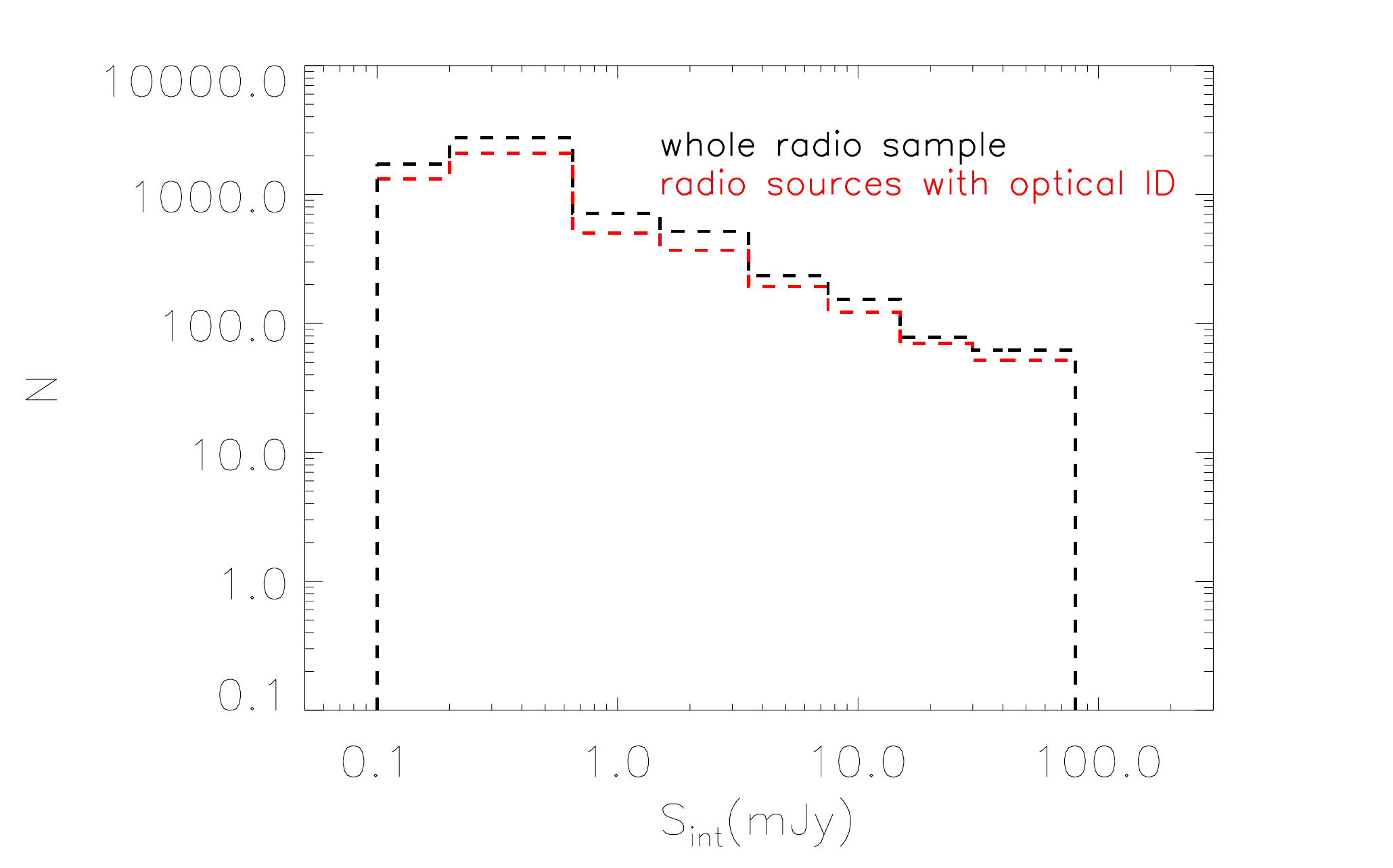} \\
        \includegraphics[width=\columnwidth]{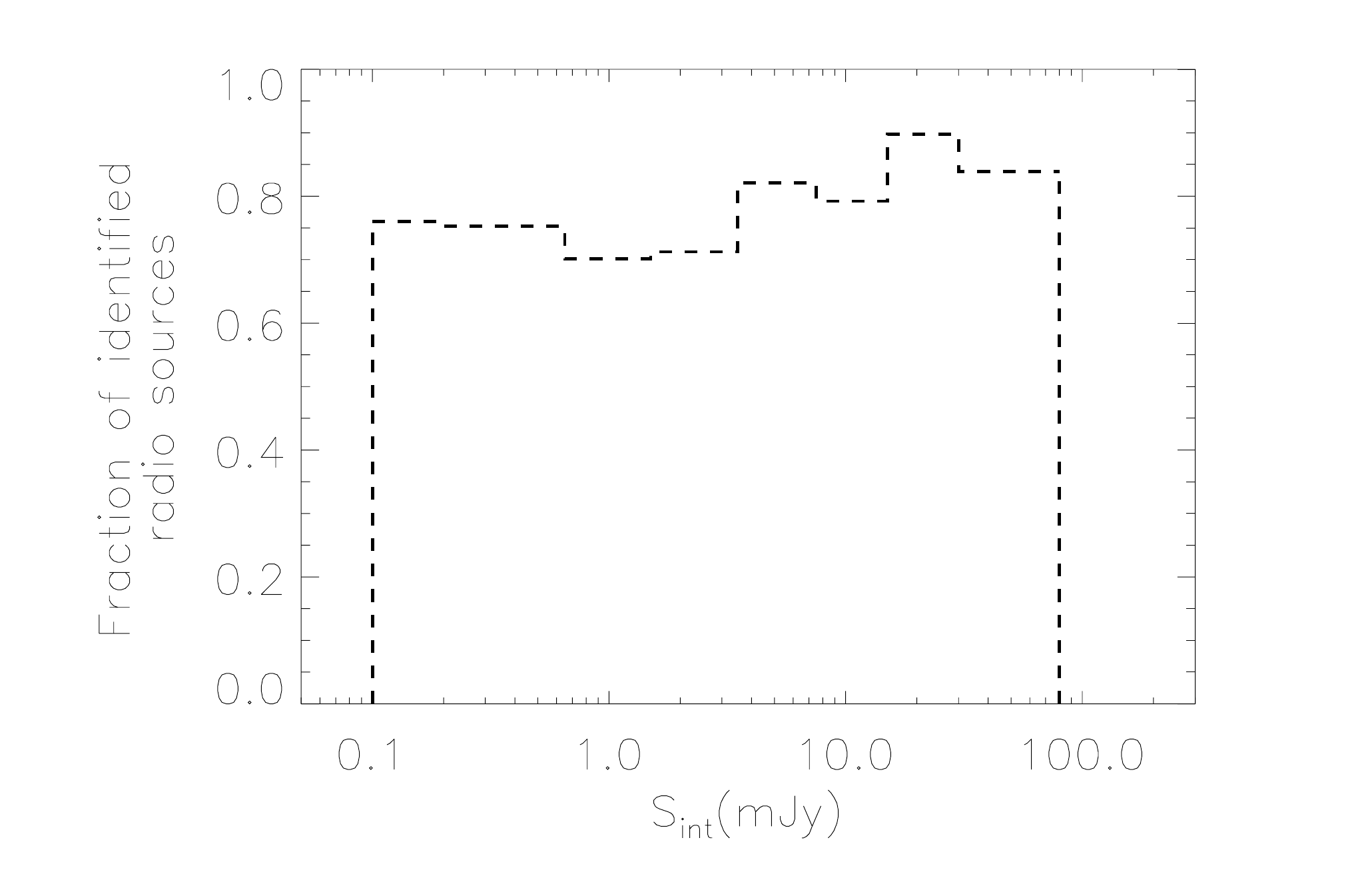} 
                        \caption{$Top$: Total radio flux distribution for the whole radio sample (black histogram) and for the 4770 radio sources with an      optical identification (red histogram). $Bottom$: 
                        Fraction of identified radio sources  as a function of the radio flux.} 
        \label{radiotot}
        
\end{figure}

\end{center}


\begin{figure} 
        \includegraphics[width=\columnwidth]{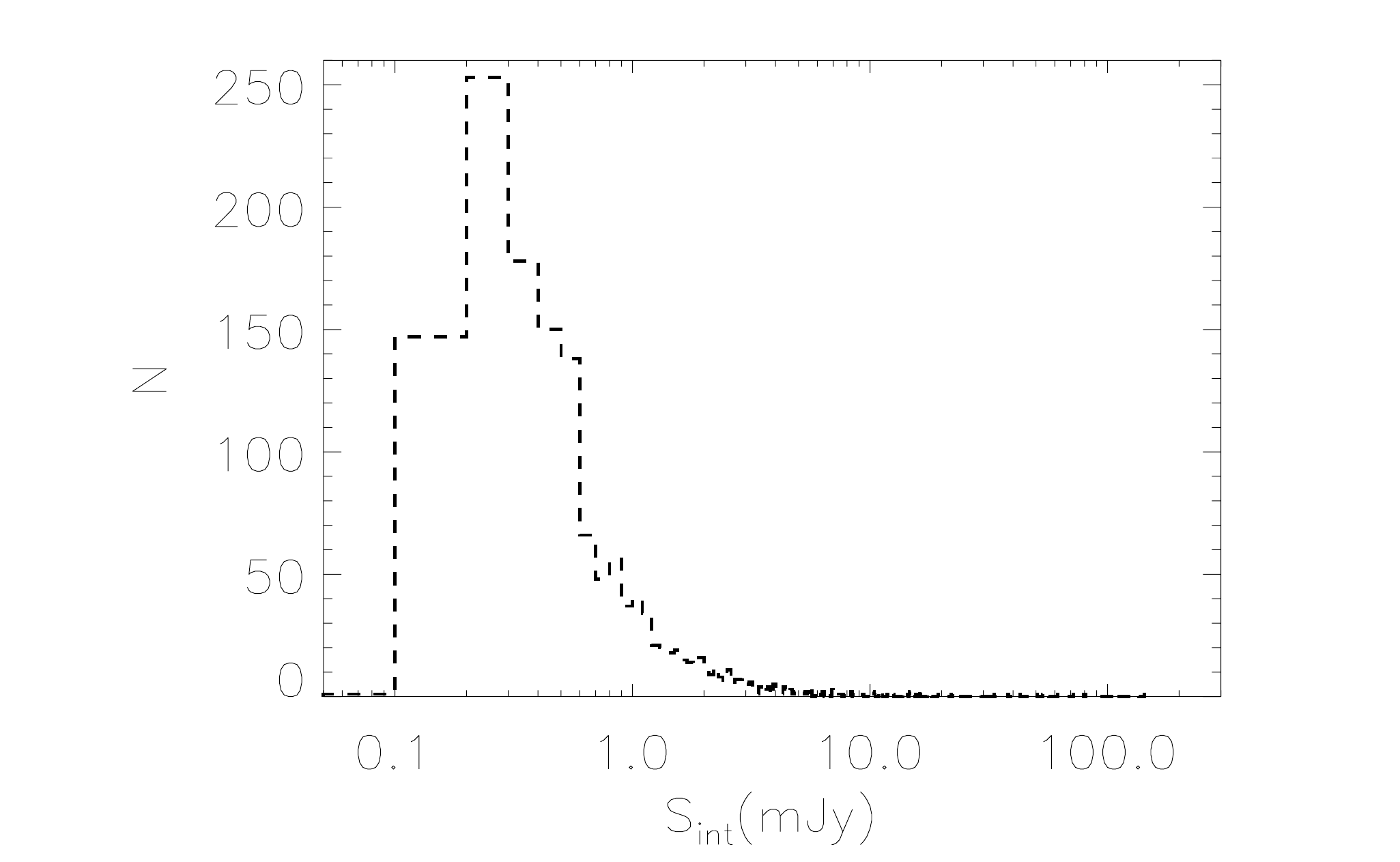} \\
        
        \caption{Total radio flux distribution for   1517 radio sources unidentified in the optical and NIR bands. }  
        \label{radiounid}       
\end{figure}

\subsection{Unidentified radio sources} 

Excluding the 111 radio sources for which no optical/NIR data are available,  we have a  total of 1406 unidentified radio sources. 

\begin{figure} 
        \includegraphics[width=\columnwidth]{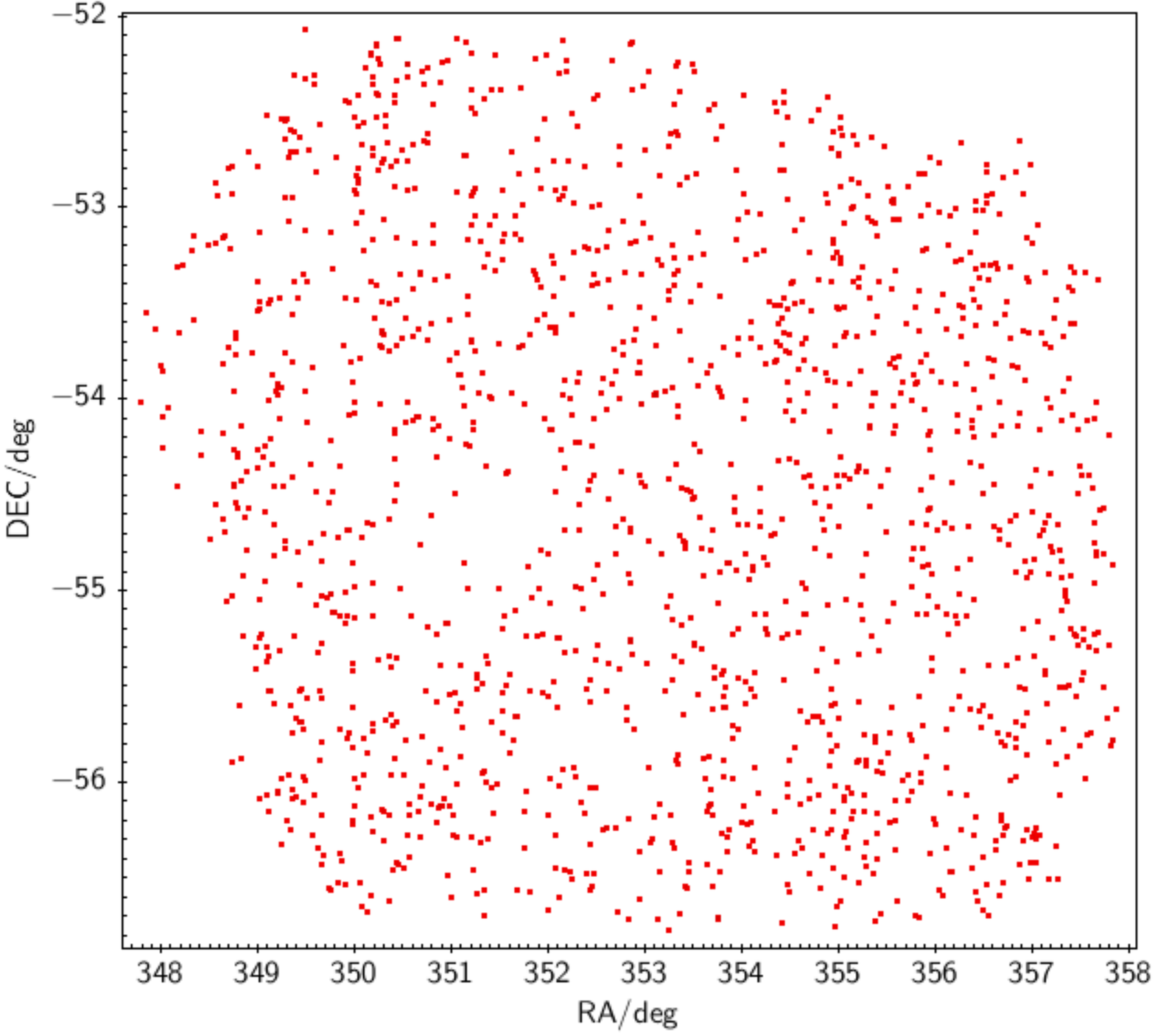} \\
                \caption{Position of the 1406 unidentified radio sources. }  
        \label{radiounidRA_DEC} 
\end{figure}

In Fig.~\ref{radiotot}  (top panel) we show the integrated radio flux distribution $S_{int}$ for the entire radio sample of 6287 radio sources (black histogram) and for the 4770 radio sources with an optical/NIR counterpart (orange histogram), while in 
the bottom panel we show the fraction of identified radio sources as function of the radio flux. 
In Fig.~\ref{radiounid}  we show the integrated radio flux distribution for the 1406 radio sources without optical/NIR counterparts.  As is evident from Figs.~\ref{radiotot} and ~\ref{radiounid},   the 
majority of the unidentified radio sources   are sources with the fainter radio fluxes. Finally in Fig. \ref{radiounidRA_DEC} we show the distribution in right ascension and declination of the 1406 unidentified radio sources. 

Although the area covered and the magnitude limit of the different surveys used are   not  uniform, 
 the unidentified radio sources are   smoothly 
 distributed over the  whole area and actually follow the  $z_{DEC}$  area coverage.  This result suggests
 that, as expected, the identification process is dominated by the data in the $z_{DEC}$ band, for which we have the best 
 data set both in terms of  area covered and magnitude limit.

\section{Optical properties of radio sources  }

 \begin{figure*}
               \includegraphics[width=6.2cm,clip]{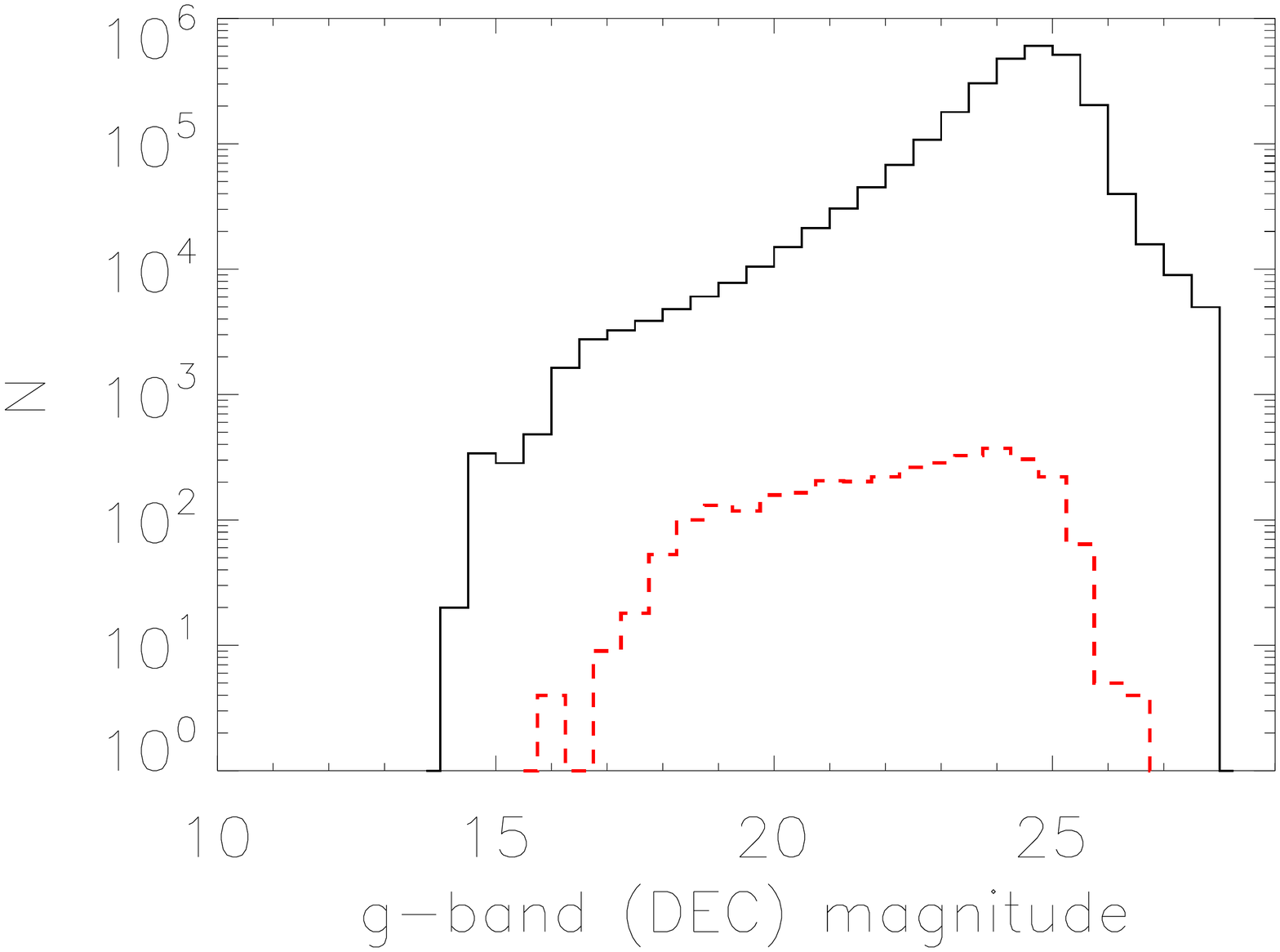}
                \includegraphics[width=6.2cm,clip]{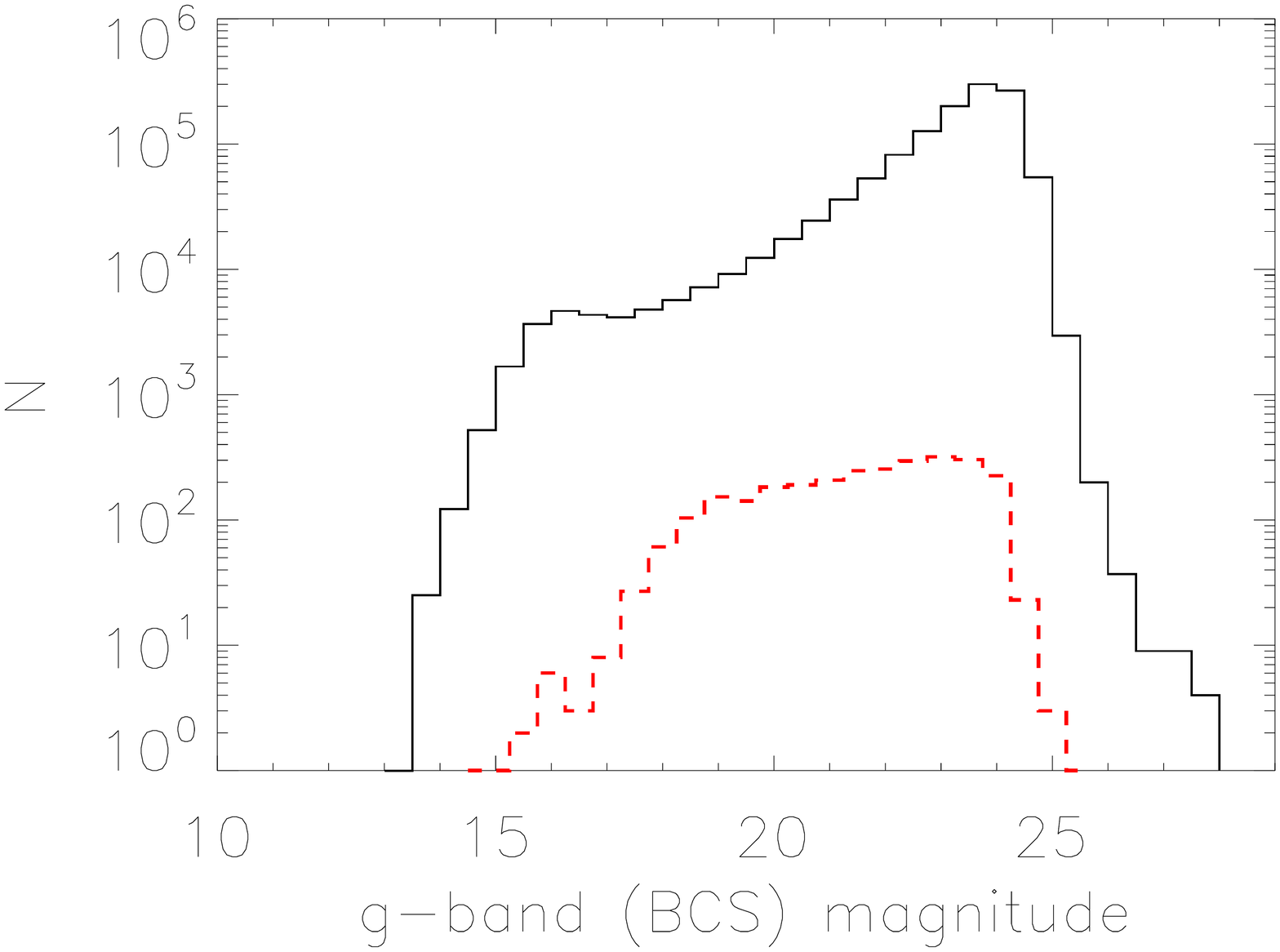}
                \includegraphics[width=6.2cm,clip]{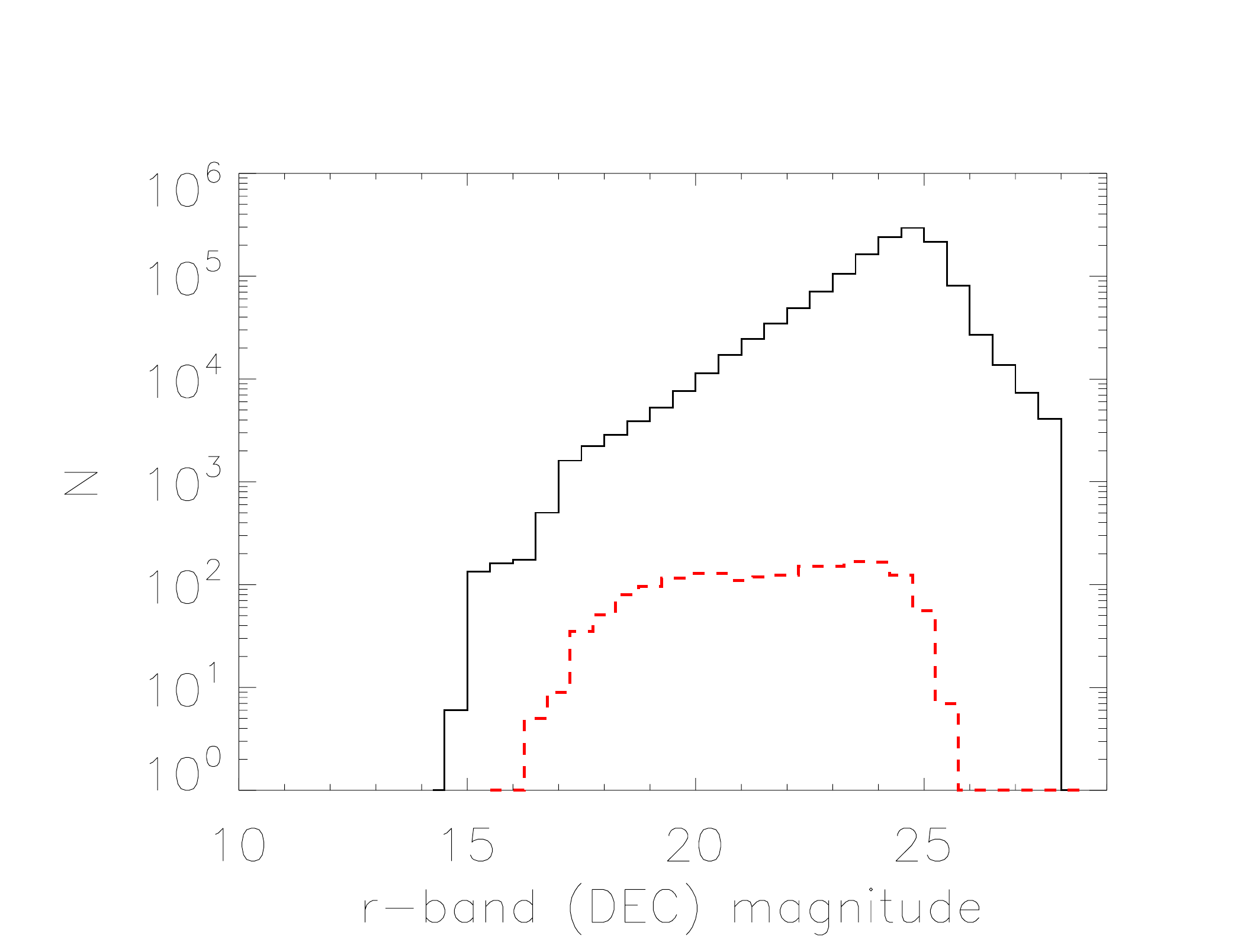}
                \includegraphics[ width=6.2cm,clip]{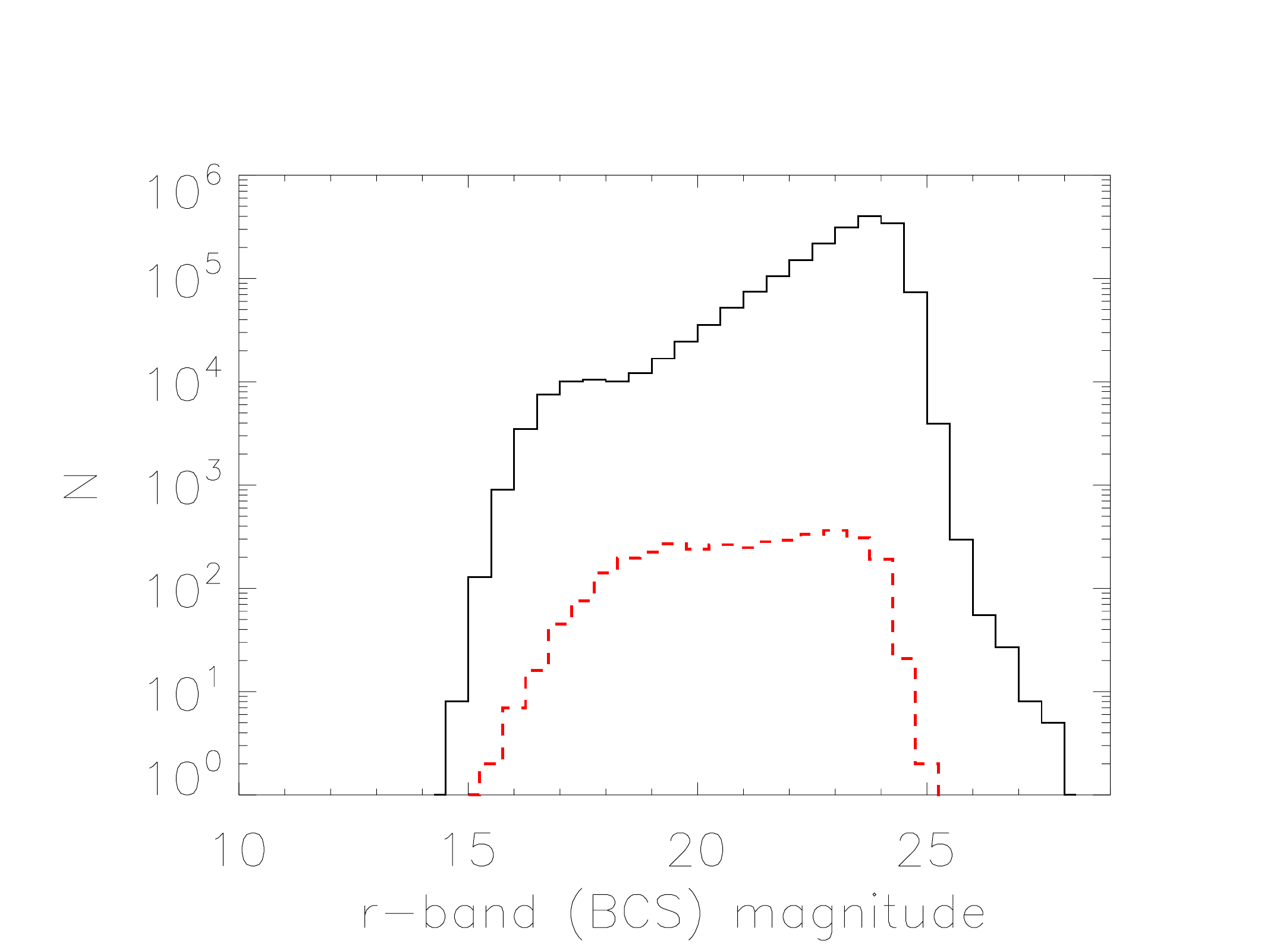}
                 \includegraphics[width=6.2cm,clip]{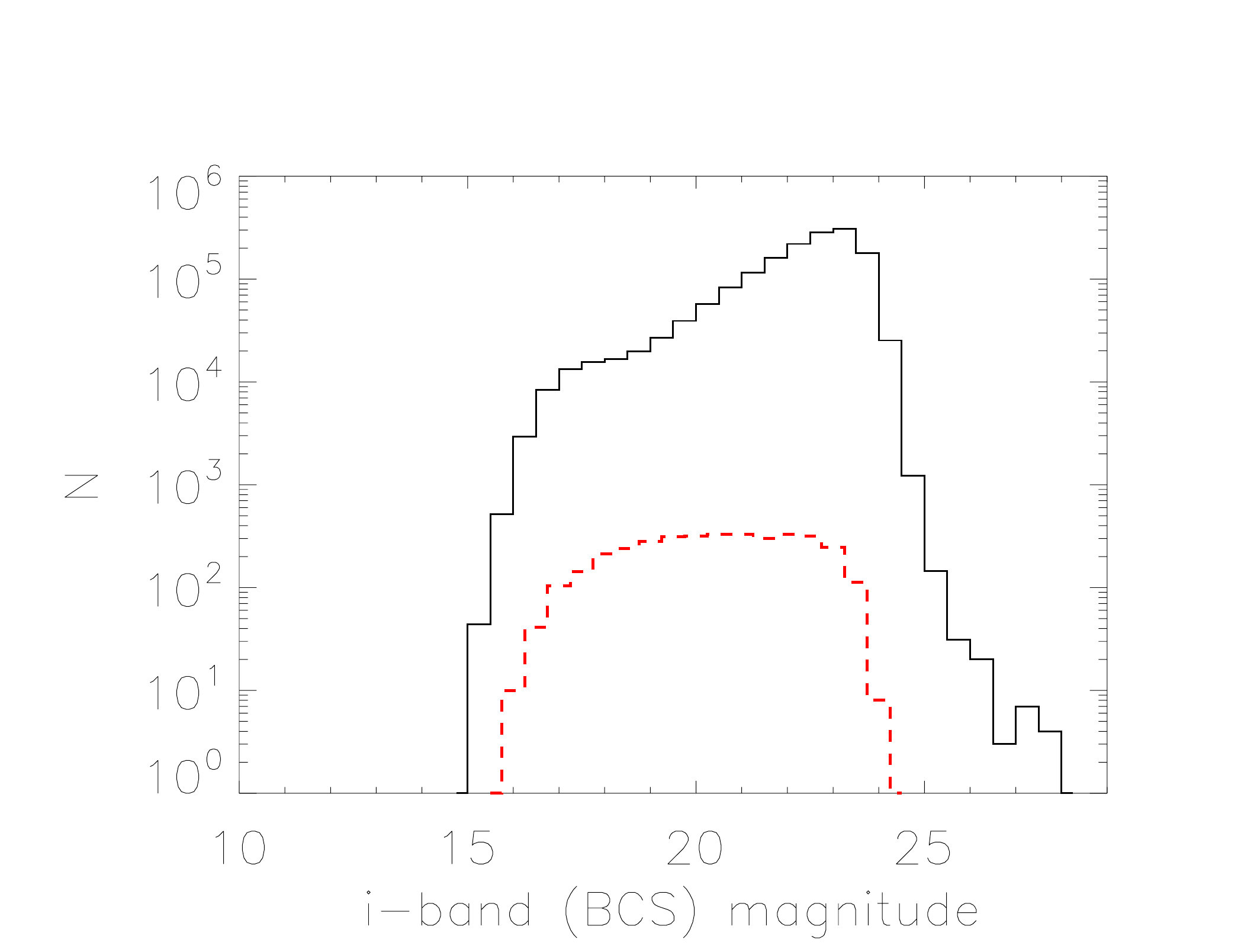}      
                 \includegraphics[width=6.2cm,clip]{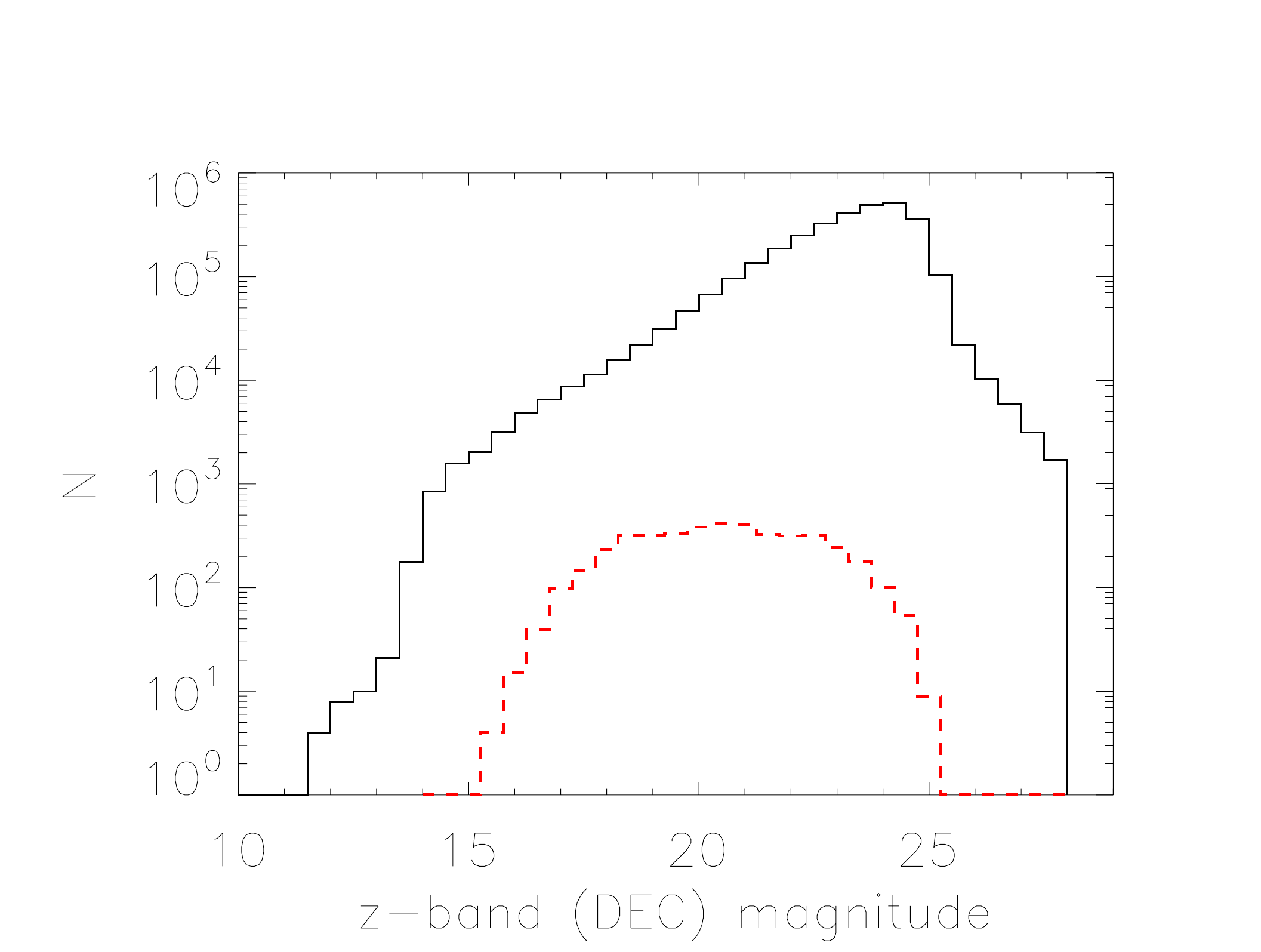}
                 \includegraphics[width=6.2cm,clip]{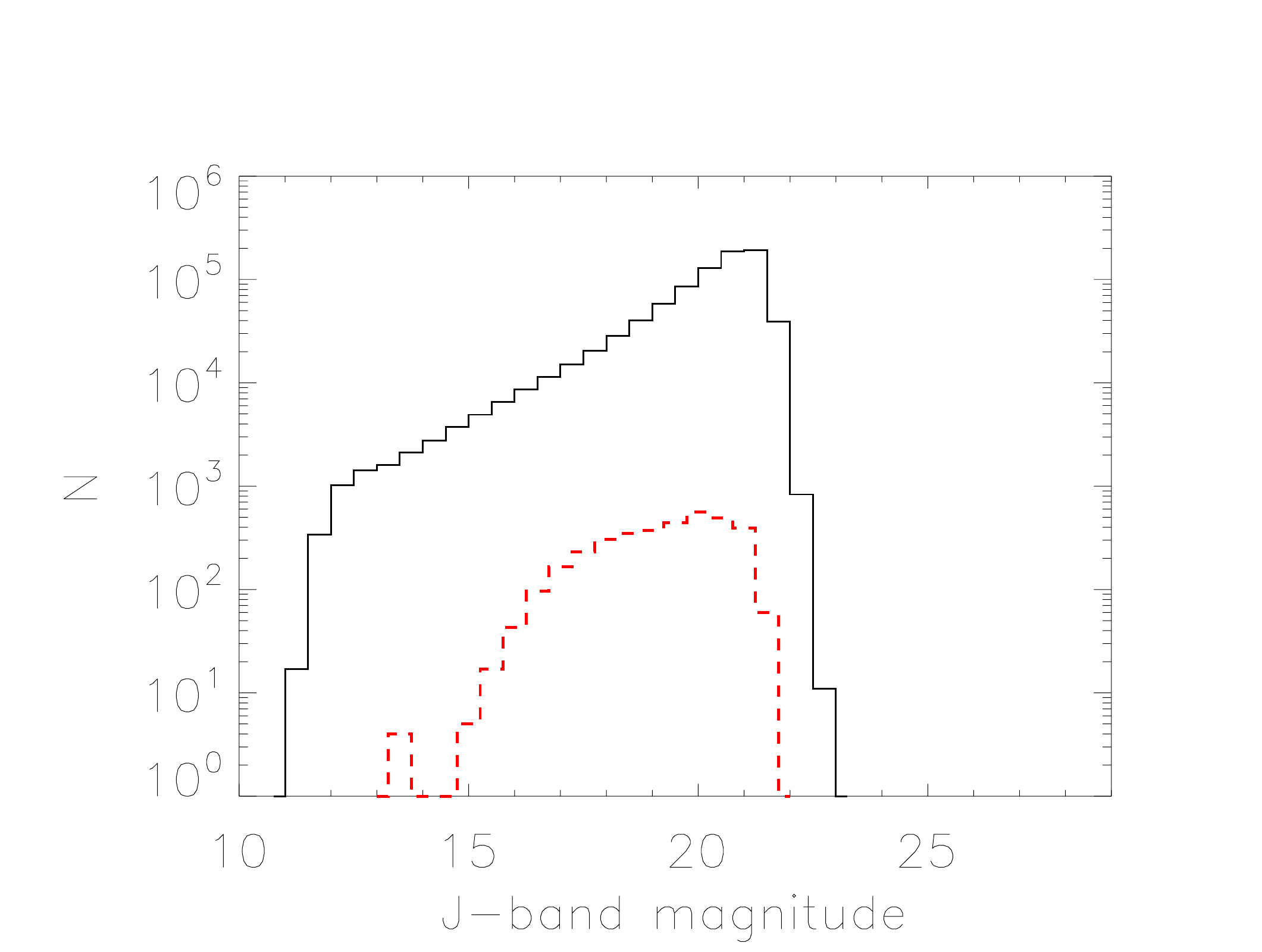}
                 \includegraphics[width=6.2cm,clip]{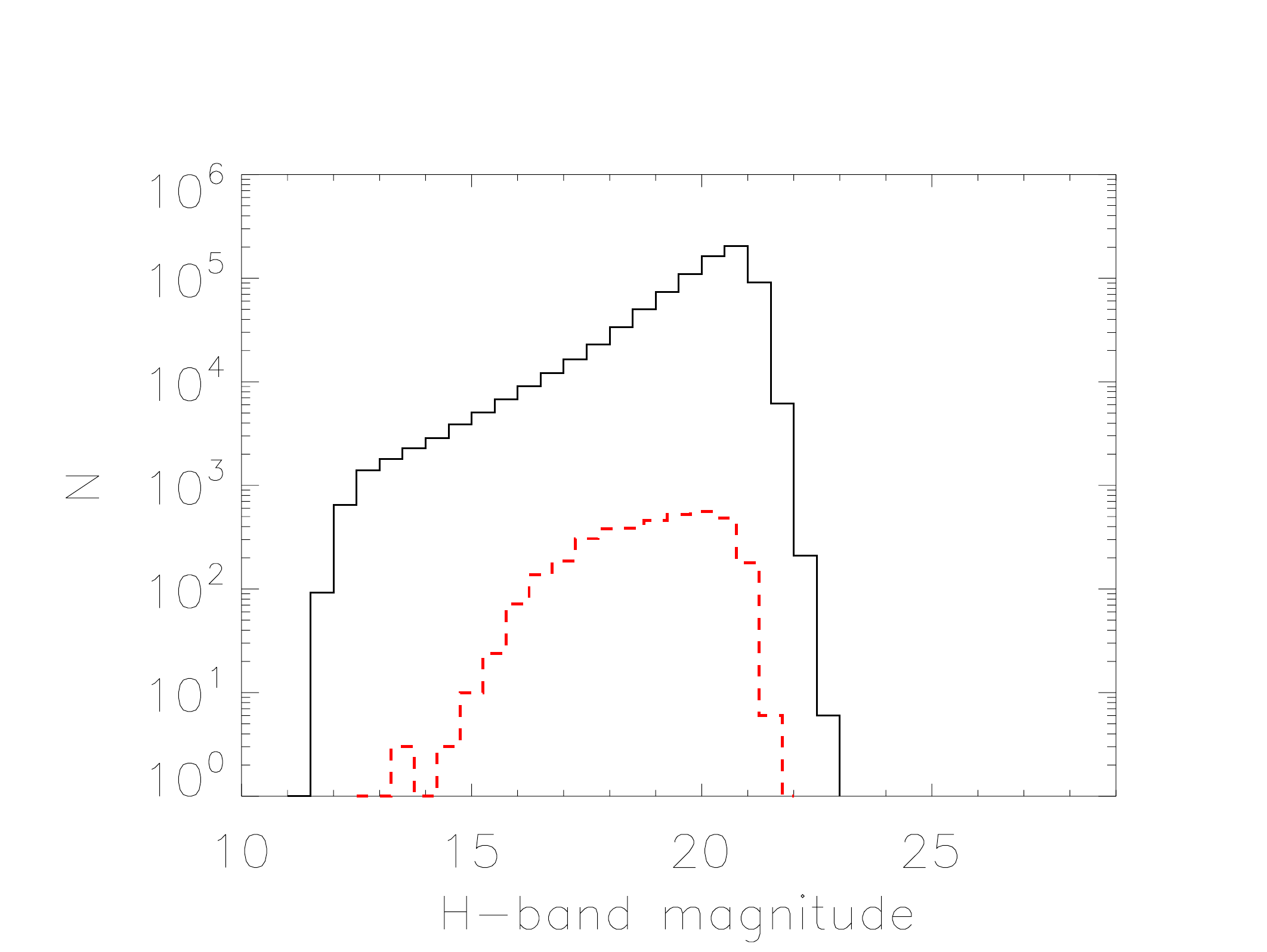}
                 \includegraphics[width=6.2cm,clip]{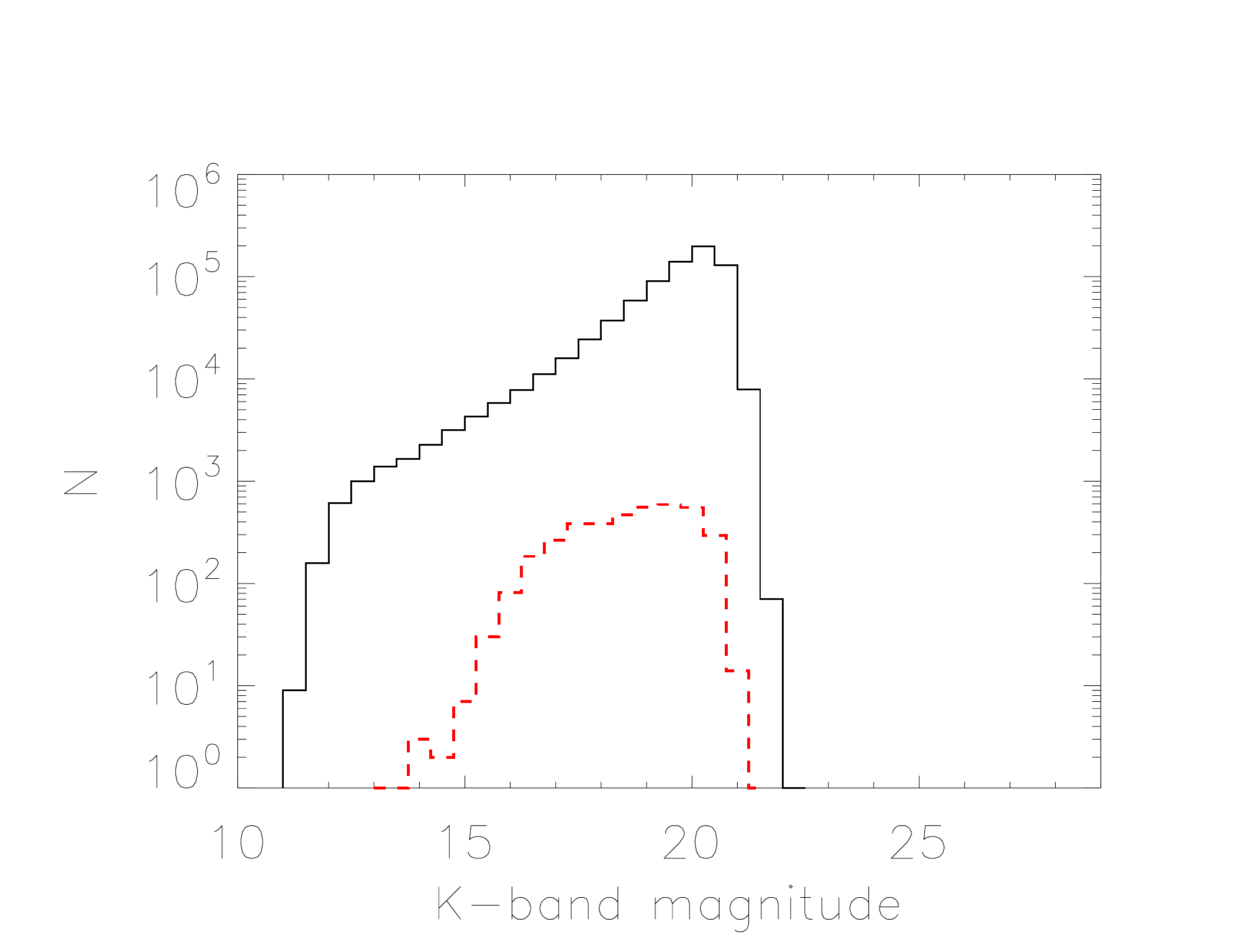}
                
                   \caption{Magnitude distributions for the entire optical/NIR  samples (solid black line) and for the optical/NIR counterparts of the 
                   radio sources (dashed red line) for all the bands considered.  } 
                   
                            \label{magdist}
   \end{figure*}

In Figure \ref{magdist} we report the magnitude distributions of the optical and NIR counterparts of the radio sources (orange dashed lines)  compared 
to the magnitude distributions of the entire XXL-S optical or NIR  samples in the relative band.    

From these figures  it is evident that the magnitude distribution of the 
radio source counterparts is flatter than that from the entire optical/NIR  sample, showing that
a significant fraction of radio sources are associated with relatively bright optical galaxies/AGN.  Since there is no correlation between the magnitude of the counterparts and the radio-optical/NIR separation 
(see bottom panel of Figure ~\ref{separation}), we can exclude that this result is due to (or at least is dominated by)   the loss of 
faint sources behind brighter sources.  

Moreover, from Figure \ref{magdist} it is interesting to note that the magnitude distributions of the
radio sources has a peak coincident with the peak of the total distribution, i.e. coincident
with the magnitude limit of the optical/NIR data. This result confirms that
the relatively low fraction of optical/NIR identifications  of the radio sources within the 
area covered by the optical/NIR data (see Table 1)  is mainly due to a relatively shallow  limiting
magnitude of the available data. 

\begin{figure*}
\centering
 \includegraphics[width=8.8cm,clip]{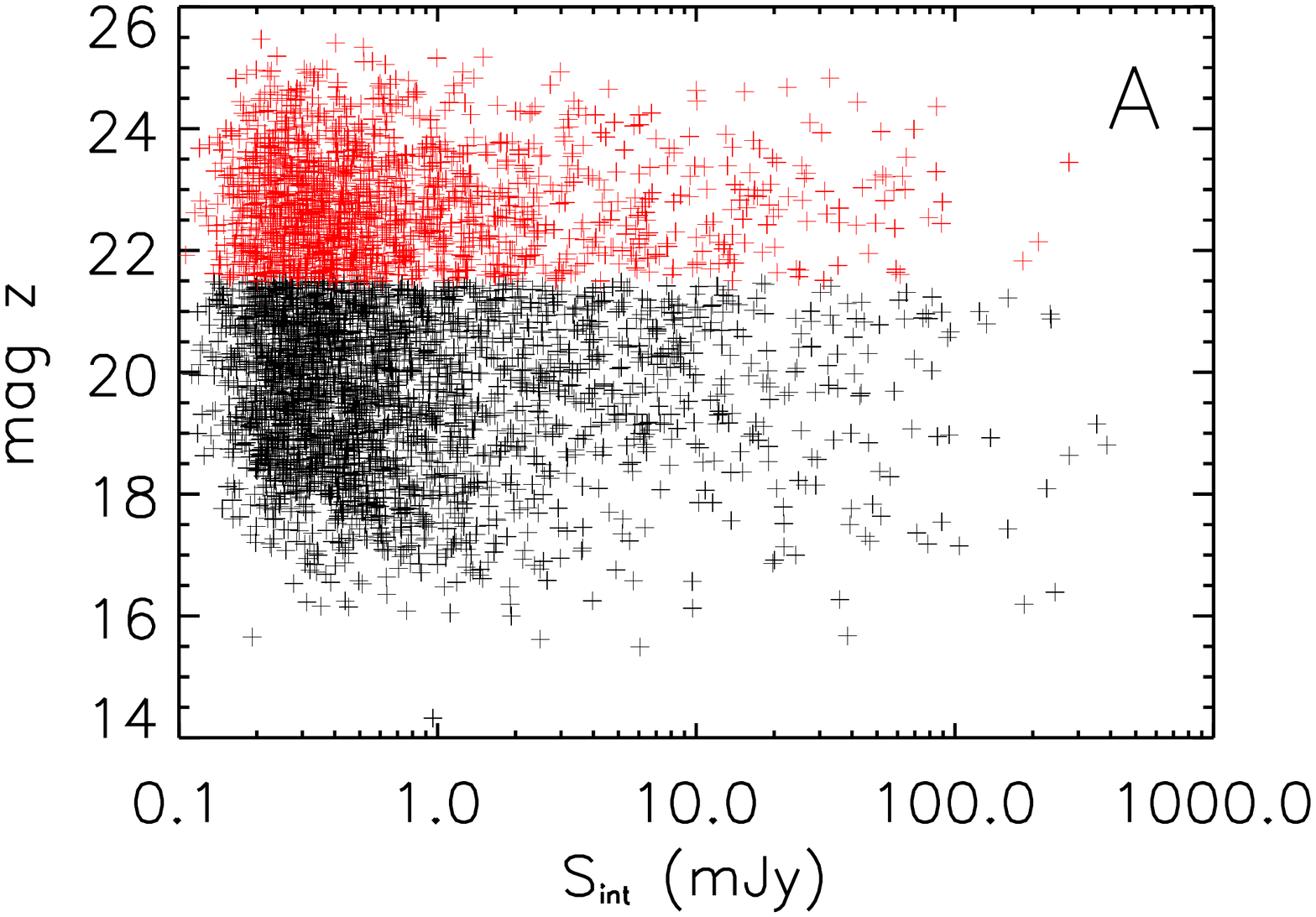}
 \includegraphics[width=8.8cm,clip]{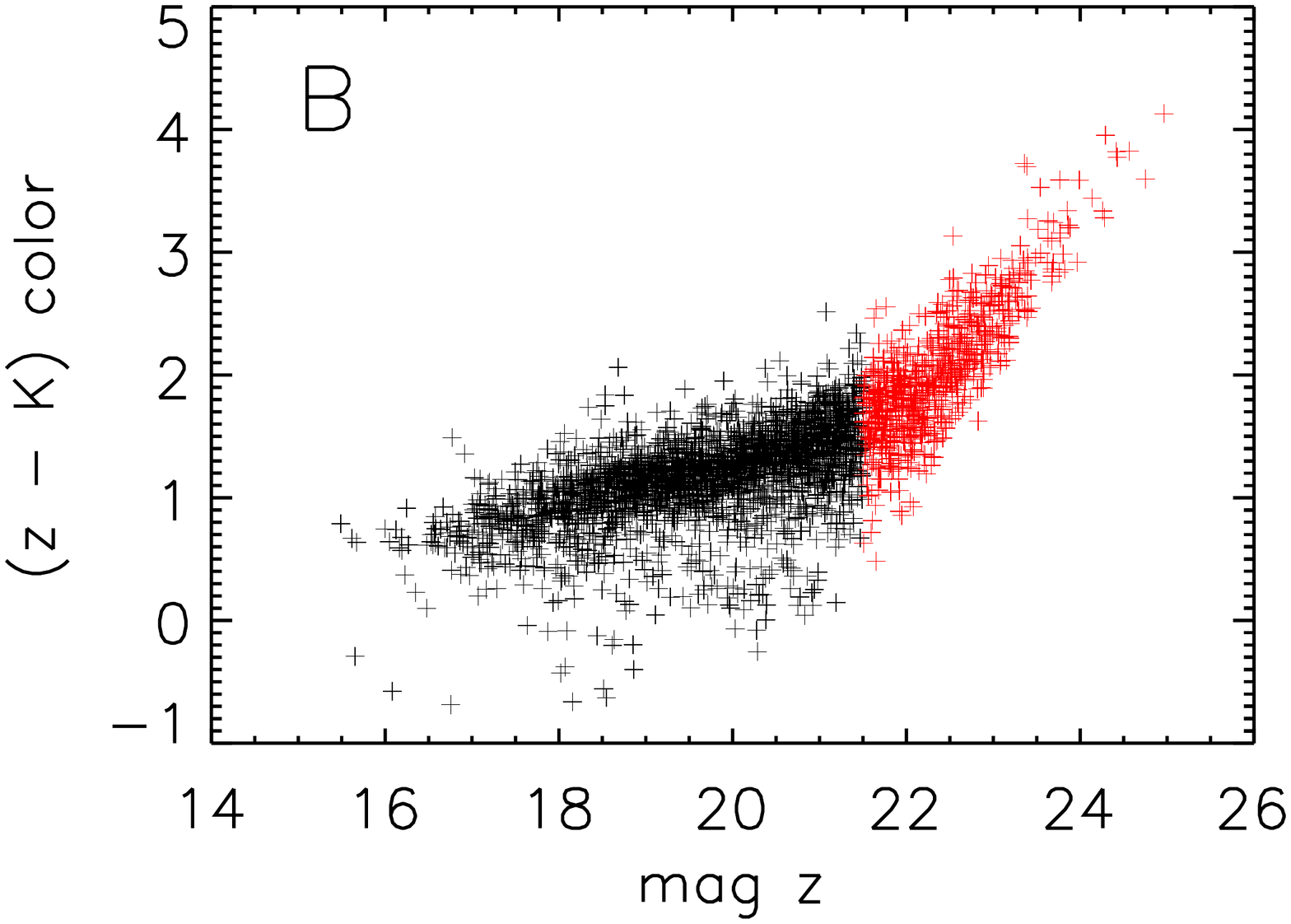}
 \includegraphics[width=8.8cm,clip]{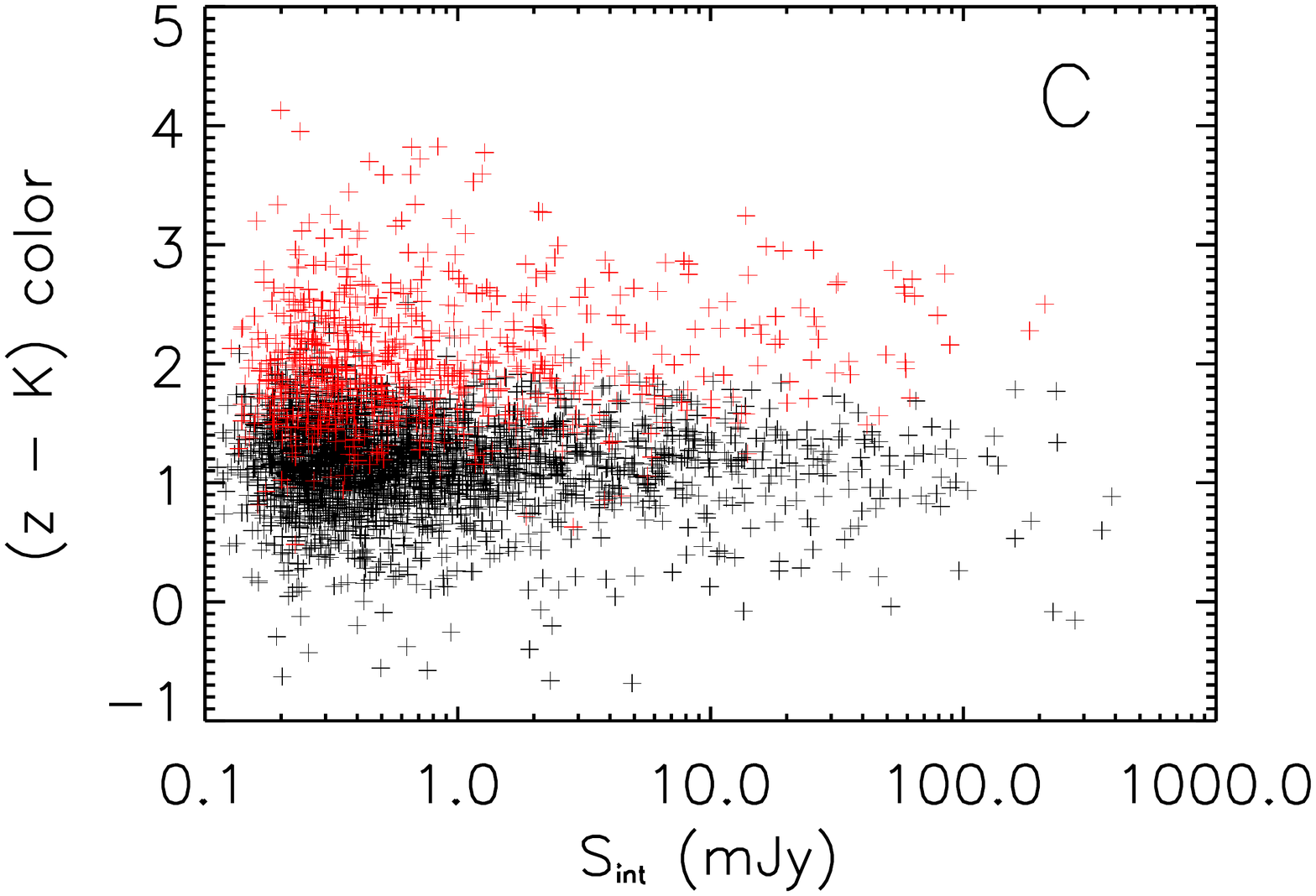}
 \includegraphics[width=8.8cm,clip]{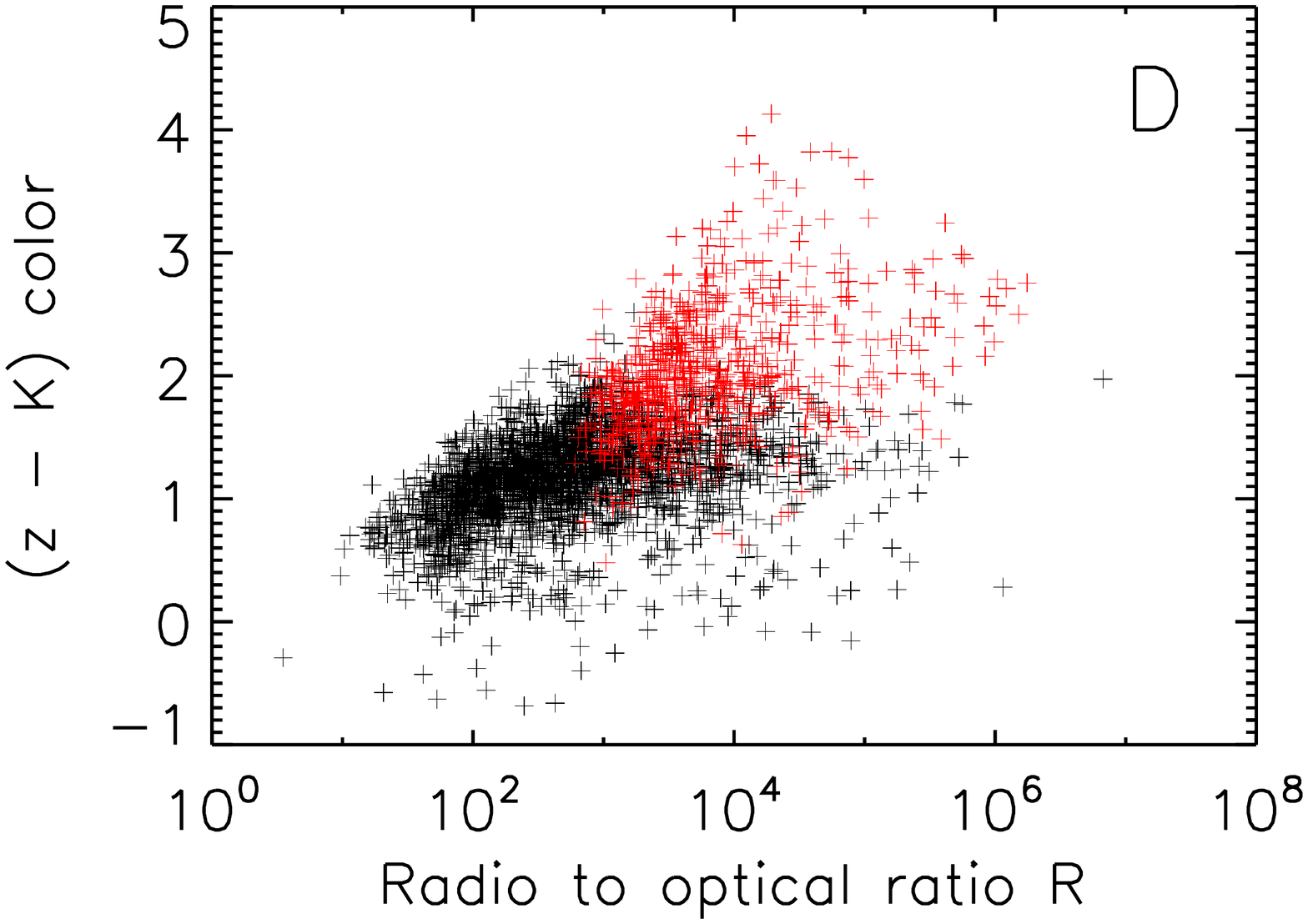}
                  \caption{Radio-optical-NIR properties of the identified radio sources. $Panel~A$:  mag $z$ from the DEC survey vs the radio flux; $Panel~B$:  ($z_{DEC}$-K) colour vs  $z_{DEC}$ magnitude; 
                      $Panel~C$:  ($z_{DEC}$-K) colour vs radio flux;   $Panel~D$:  ($z_{DEC}$-K) colour vs radio-to-optical ratio R.  Red symbols show objects with mag $z_{DEC}>$21.5 (see text for  details).  } 
                   
                            \label{mag_color}
   \end{figure*}

An overview of the optical/NIR properties of the identified radio sources is shown in the four panels of  Fig.~\ref{mag_color}, where we report mag $z_{DEC}$ versus  radio flux (Panel A); ($z_{DEC}-$K) colour versus  $z_{DEC}$ magnitude (Panel B);  ($z_{DEC}-$K) colour versus radio flux (Panel C); and  ($z_{DEC}-$K) colour versus the radio-to-optical ratio $R$ defined as 
$R=S\times10^{0.4(mag-12.5)}$, where $S$ and $mag$ are the radio flux in mJy and the apparent magnitude of the sources, respectively  (Panel D).  The red symbols in all panels show objects with mag $z_{DEC}>$21.5.   From the figure  it is evident that while there is no correlation between the magnitude and the radio flux (Panel A),  redder objects (with $z_{DEC}-$K) $>$ 2) become evident only in optically deep 
surveys, i.e.  at   magnitude  fainter than $z\sim$21.5 (Panel B), although present  at any radio flux levels (Panel C).  Finally, Panel D shows that redder object (with ($z_{DEC}-$K) $>$ 2)  are only associated  
with radio sources with an optical-to-radio ratio $R>$1000, i.e.  sources generally associated with early-type galaxies whose radio emission is dominated by nuclear activity (Ciliegi et al. 2003). 

Figure~\ref{mag_color} shows that regardless of the radio flux limit of a radio survey, the nature of the identified sources  is strongly dependent   on the depth of the optical/NIR used in the identification 
process.  As first noted by Gruppioni et al. (1999), only with deep enough optical/NIR data  will we be able to identify a significant fraction of radio sources with red ($z_{DEC}-$K)  counterparts, whose radio emission  is dominated by nuclear activity rather than 
starburst activity. This result must be taken into account in any identification processes of radio sources and can explain the difference in the results of the radio sources identification  
found in different works (see discussion in  Section 1 and reference therein). 

The classification of the 4770 optical/NIR counterparts of the radio sources are beyond the scope of this paper, but they are described in XXL Paper XXXI.  The results are described briefly as follows.
 Using several different diagnostic diagrams based on 
the photometric properties (from radio to X-ray) of the sources, as well as optical spectral templates, 
emission lines, and colours,  XXL Paper XXXI classified the 4770 optical/NIR counterparts
as   12 stars,  1729 low-excitation radio galaxies (LERGs), 1159 radio-loud high-excitation radio galaxies (RL HERGs), 296 radio-quiet 
high-excitation radio galaxies (RQ HERGs),  558 star-forming 
galaxies (SFGs), and 1016 unclassified sources.  
For a detailed description of the classification and properties of the cross-matched radio sources, see XXL Paper XXXI.

\section{Conclusion}

Starting from the sample of 6287  radio sources detected at 2.1 GHz  in the XXL-S field (XXL Paper XVIII) 
and from the multiwavelength catalogue (from optical to X-ray bands)  available in the same field 
(XXL Paper VI) we used the likelihood ratio technique to identify the  counterparts of the radio 
sources.   Since the optical photometric data cover slightly different areas of the sky, in order to maximise the number of identifications 
we applied this technique  to all the available data:  seven optical ($g_{BCS}$, $g_{DEC}$, $r_{BCS}$, $r_{DEC}$, $i_{BCS}$, $i_{DEC}$, $z_{DEC}$) and three NIR ($J$, $H$, $K$) bands.  We found optical/NIR counterparts for  4770 different radio sources in the XXL-S field, 414 of which have also been detected  in the X-ray band.         

This fraction of identification  is in agreement with  previous radio-optical association studies at similar optical magnitude depth (Ciliegi et al. 2005),  although  is relatively low in comparison to recent  work conducted in other radio fields using deeper optical and NIR data (Smol{\v c}i{\'c} et al. 2017a).  The analysis of the magnitude distributions of the optical/NIR samples and of the counterparts of the radio 
sources shows that the magnitude distributions of the
radio sources has a peak coincident with the peak of the total distribution, i.e. coincident
with the magnitude limit of the optical/NIR data.  This confirms that with deeper  optical/NIR data we could expect
an optical magnitude distribution of the radio
sources with a maximum brighter than the limiting magnitude and then a higher fraction of identifications.  \\
Finally, the analysis of the  optical/NIR properties of the identified 
radio sources shows  that the depth of the data used in the identification process plays a significant role in establishing the nature of the identified radio sources. 

The sample of the 4770 counterparts of the radio sources has been analysed and classified  in XXL Paper XXXI.    
Further papers to  fully explore the properties of the radio sources 
in the XXL-S field are in progress within the XXL collaboration. 

\section{Acknowledgements}

XXL is an international project based around an XMM Very Large Programme surveying two 25 deg$^2$ extragalactic fields at
a depth of $\sim$ 6 $\times$ 10 $^{-15}$ erg cm $^{-2}$ s${-1}$  in the [0.5--2.0] keV band
for point-like sources. The XXL website is http://irfu.cea.fr/xxl. Multi-band information and spectroscopic follow-up of the X-ray
sources are obtained through a number of survey programmes, summarised at http://xxlmultiwave.pbworks.com/. The Australia
Telescope Compact Array is part of the Australia Telescope National Facility, which is funded by the Australian Government for
operation as a National Facility managed by CSIRO. Based in part on data acquired through the Australian Astronomical Observatory
via programme A/2016B/107.   SF acknowledges support from  the  Swiss  National  Science  Foundation. 
VS acknowledges support from the European Union's Seventh Frame-work programme under grant agreement 333654 (CIG, `AGN feedback').
JD acknowledges support from the European Unions Seventh
Frame-work programme under grant agreement 337 595 (ERC Starting Grant, `CoSMass').
The Saclay group acknowledges long-term support from the Centre National d'Etudes Spatiales (CNES).





\begin{thebibliography}{}

\bibitem{} Afonso, J., Georgakakis, A., Almeida, C., Hopkins, A.M., Cram, L.E., Mobasher, B., Sullivan, M.,  2005,  ApJ, 624, 135 

\bibitem{}  Ashby, M.L.N., Stanford, S.A., Brodwin, M., et al.  2013, ApJS, 209, 22 

\bibitem{} Bardelli, S., Schinnerer, E.,  Smol{\v c}i{\'c}, V.,  et al. 2010, A\&A, 511, 1 

\bibitem{} Bondi, M.,  Ciliegi, P., Zamorani, G., et al.,  2003, A\&A,  403, 857 

\bibitem{} Bondi, M.,  Ciliegi, P., Schinnerer, E., Smol{\v c}i{\'c}, V., Jahnke, K.,  Carilli, C., Zamorani, G., 2008, ApJ, 681, 1129

\bibitem{} Brusa, M.,  Zamorani, G., Comastri, A.,  et al.  2007,  ApJSS, 172, 353  

\bibitem {}Butler, A., Huynh, M., Delheize, J., et al. 2017,  A\&A in press, arXiv170310296, (XXL Survey, Paper XVIII) 

\bibitem{}Butler, A., Huynh, M., Delheize, J., et al. 2018, A\&A, in press, arXiv1804.05983, (XXL Survey,  Paper XXXI) 

\bibitem {}Carilli et al. 2007, ApJS, 172, 518

\bibitem{} Chiappetti, L., et al., 2018, in preparation, (XXL Survey, Paper XXVII) 

\bibitem {}Ciliegi, P., Zamorani G., Hasinger G., Lehmann I., Szokoly G.,  Wilson G., 2003,  A\&A,  398, 901

\bibitem{}Ciliegi, P.,  Zamorani, G., Bondi, M., et al.  2005,  A\&A, 441,  879

\bibitem{} Condon, J.J., 1984, ApJ, 287, 461

\bibitem{} Condon, J.J., 1992, ARAA, 30, 575

\bibitem{} Condon, J.J., Cotton, W.D. and Broderick, J.J., 2002, ApJ, 124, 675

\bibitem{} Danese, L.,  De Zotti, G., Franceschini, A. and Toffolatti, L.,  1987, ApJ, 318, L15

\bibitem{} Delhaize, J., Smol{\v c}i{\'c}, V.,  Delvecchio, I., et al., 2017, A\&A, 602, 4

\bibitem {}Delvecchio, I., Smol{\v c}i{\'c}, V., Zamorani, G.,    et al. 2017,  A\&A, 602,  A3

\bibitem{}Desai, S., Armstrong, R., Mohr, J.J., et al. 2012, ApJ, 757, 83

\bibitem{} Dubner, G., and Giacani, E., 2015, A\&AR, 23, 3 

\bibitem{} Fomalont, E.B.,  Kellermanm, K.I., Cowie, L.L., et al. 2006, ApJS,  167, 103 

\bibitem{}Fotopoulou, S, Pacaud F., Paltani S. et al. 2016,   A\&A,  592, 5, (XXL Survey, Paper  VI)

\bibitem{} Gruppioni, C., Mignoli, M. and Zamorani, G.,  1999,  304, 199

\bibitem{} Hammer, F.,  Crampton, D., Lilly, S., LeFevre, O., and Kenet, T.,  1995, MNRAS, 276, 1085 

\bibitem{} Huynh, M.T.,,   Jackson, C.,  Norris, R., and Prandoni, I., 2005,  AJ, 130, 1373 

\bibitem{} Martin, C. \& GALEX Team 2005, in IAU Symposium, Vol 216, Maps of the Cosmos, ed. M. Colles, L. Stavely-Smith, \& R.A. Stathakis, 221 

\bibitem{} Morrissey, P., Schiminovich, D.,  Barlow T.A., et al., 2005, ApJ, 770, 40

\bibitem{} Novak, M., Smol{\v c}i{\'c}, V., Delhaize, J., et al., 2017, A\&A, 602, 5

\bibitem{} Padovani, P.,  Manieri, V., Tozzi, P., Kellermann, K.I., Fomalont, E.B., Miller, N., Rosati, P., and Shaver, P., 2007, ASPC, 380, 205

\bibitem{} Padovani, P., Manieri, V., Tozzi, P., Kellermann, K.I., Fomalont, E.B., Miller, N., Rosati, P., and Shaver, P., 2009, ApJ, 694, 235

\bibitem{} Pierre, M., Pacaud, F., Adami, C., et al. 2016, A\&A, 592, A1, (XXL Survey,  Paper I) 

\bibitem{} Prandoni, I., Gregorini, L., Parma, P., de Ruiter, H.R., Vettolani, G., Zanichelli, A., Wieringa, M.H., Ekers, R.D.,  2001, A\&A, 369, 787 

\bibitem{} Richards, E.A., 2000, ApJ, 533, 611

\bibitem{} Sadler, E.M., McIntyre, V.J., Jackson, C.A. and Cannon, R.D., 1999, PASP, 16, 247

\bibitem{}Schinnerer, E., Smol{\v c}i{\'c}, V.,, Carilli, C., et al.,  2007, ApJS, 172, 46

\bibitem{}Schinnerer, E., Sargent, M.T., Bondi, M., et al. 2010, ApJS,  188, 384

\bibitem{} Scoville, N., Aussel, H.,  Brusa, M., et al., 2007, ApJS, 172, 1

\bibitem {}Smol{\v c}i{\'c}, V.,  Schinnerer, E., Scodeggio, M., et al., 2008, ApJSS, 177, 14

\bibitem {}Smol{\v c}i{\'c}, V., Zamorani, G., Schinnerer, E., et al., 2009a,  ApJ, 696, 24 

\bibitem {}Smol{\v c}i{\'c}, V., Schinnerer, E., Zamorani, G., et al., 2009b, ApJ, 690, 610

\bibitem{} Smol{\v c}i{\'c}, V., Delhaize, J., Huynh, M.,  et al. 2016, A\&A, 592, 10, (XXL Survey, Paper XI)

\bibitem {}Smol{\v c}i{\'c}, V., Delvecchio, I., Zamorani, G., et al. 2017a, A\&A, 602, A2 

\bibitem {}Smol{\v c}i{\'c}, V.,  Novak, M., Bondi, M., et al, 2017b, A\&A,  602, A1 

\bibitem{} Sutherland, W.,   and  Saunders, W.,  1992,  MNRAS, 259, 413 

\bibitem{} White, R.L., Becker. R.H., Helfand, D.J., Gregg, M.D., 1997, ApJ, 475, 479 

\bibitem{} Windhorst, R.A., Miley. G.K., Owen, F.N., Kron, R.G., and Koo, D.C., 1985 , ApJ, 289, 494

\bibitem{} Wright, E.L., Eisenhardt, P.R.M., Mainzer, A.K., et al, 2010, AJ, 140, 1868

\end{thebibliography}
\end{document}